\documentclass[aps,prl,twocolumn,showpacs,floatfix,nobibnotes,nofootinbib,superscriptaddress,amsmath,amssymb,longbibliography]{revtex4-1}

\usepackage{graphicx}
\usepackage{epsfig}
\usepackage{bm}
\usepackage{color}
\usepackage{float}
\usepackage{dcolumn}
\usepackage{multirow} 
\usepackage{hyperref}
\usepackage{lipsum}

\newcommand{\beq}{\begin{equation}}
\newcommand{\eeq}{\end{equation}}
\newcommand{\beqa}{\begin{eqnarray}}
\newcommand{\eeqa}{\end{eqnarray}}

\begin{document}

\title{Global sensitivity analysis of bulk properties of an atomic nucleus}

\author{Andreas~Ekstr\"om} \affiliation{Department of Physics,
  Chalmers University of Technology, SE-412 96 G\"oteborg, Sweden}

\author{Gaute~Hagen}
\affiliation{Physics Division, Oak Ridge National Laboratory,
Oak Ridge, TN 37831, USA}
\affiliation{Department of Physics and Astronomy, University of Tennessee,
Knoxville, TN 37996, USA}

\date{\today}

\begin{abstract} 
We perform a global sensitivity analysis of the binding energy and the
charge radius of the nucleus $^{16}$O to identify the most influential
low-energy constants in the next-to-next-to-leading order chiral
Hamiltonian with two- and three-nucleon forces. For this purpose we
develop a subspace-projected coupled-cluster method using eigenvector
continuation [Frame D. {\it et al.}, Phys. Rev. Lett. 121, 032501
  (2018)].  With this method we compute the binding energy and charge
radius of $^{16}$O at more than one million different values of the 16
low-energy constants in one hour on a standard laptop. For relatively
small subspace projections, the root-mean-square error is about 1\%
compared to full space coupled-cluster results.  We find that 58(1)\%
of the variance in the energy can be apportioned to a single
contact-term in the $^3S_1$-wave, whereas the radius depends
sensitively on several low-energy constants and their higher-order
correlations. The results identify the most important parameters for
describing nuclear saturation, and help prioritize efforts for
uncertainty reduction of theoretical predictions. The achieved
acceleration opens up for an array of computational statistics
analyses of the underlying description of the strong nuclear
interaction in nuclei across the Segr\'e chart.
\end{abstract}


\maketitle

{\it Introduction.}- How do properties of atomic nuclei depend on the
underlying interaction between protons and neutrons? Recent \textit{ab
  initio} computations of
nuclei~\cite{hergert2014,elhatisari2015,rosenbusch2015,hagen2015,garciaruiz2016,lapoux2016,hagen2016b,simonis2017,duguet2017,morris2017,lu2019,liu2019,taniuchi2019,Gysbers2019,holt2019,soma2019}
have revealed that observables such as binding energies, radii,
spectra, and decay probabilities are very sensitive to the values of
the low-energy constants (LECs) in chiral Hamiltonian models with two-
and three-nucleon
forces~\cite{vankolck1999,epelbaum2009,machleidt2011}. Certain
interaction models work better than others, but only for selected
types of observables and in limited regions of the Segr\'e chart. It
is not clear why. The NNLO$_{\rm sat}$ interaction~\cite{ekstrom2015}
reproduces experimental binding energies and charge radii for nuclei
up to mass $A \sim
50$~\cite{hagen2015,garciaruiz2016,duguet2017,soma2019}, while the
1.8/2.0 (EM) interaction~\cite{nogga2004,hebeler2011} reproduces
binding energies and low-lying energy spectra up to mass $A \sim
100$~\cite{hagen2015,hagen2016b,morris2017,simonis2017,liu2019,holt2019}
while radii are underestimated.

To improve theoretical predictions requires rigorous uncertainty
quantification and sensitivity analyses that are grounded in the
description of the underlying nuclear Hamiltonian. Unfortunately, the
number of model samples required for statistical computing increases
exponentially with the number of uncertain LECs. A global sensitivity
analysis of the ground-state energy and charge radius $^{16}$O, based
on a realistic next-to-next-to-leading order (NNLO) chiral Hamiltonian
with $16$ LECs, requires more than one million model
evaluations. Similar numbers can be expected for Markov Chain Monte
Carlo sampling of Bayesian marginalization end evidence
integrals~\cite{Schindler2009,
  Wesolowski2015,Wesolowski2019}. Clearly, this is not feasible given
the computational cost of existing state-of-the-art \textit{ab initio}
many-body methods applied to medium-mass and heavy nuclei.
\begin{figure}[hbt]
  \includegraphics[width=0.9\columnwidth]{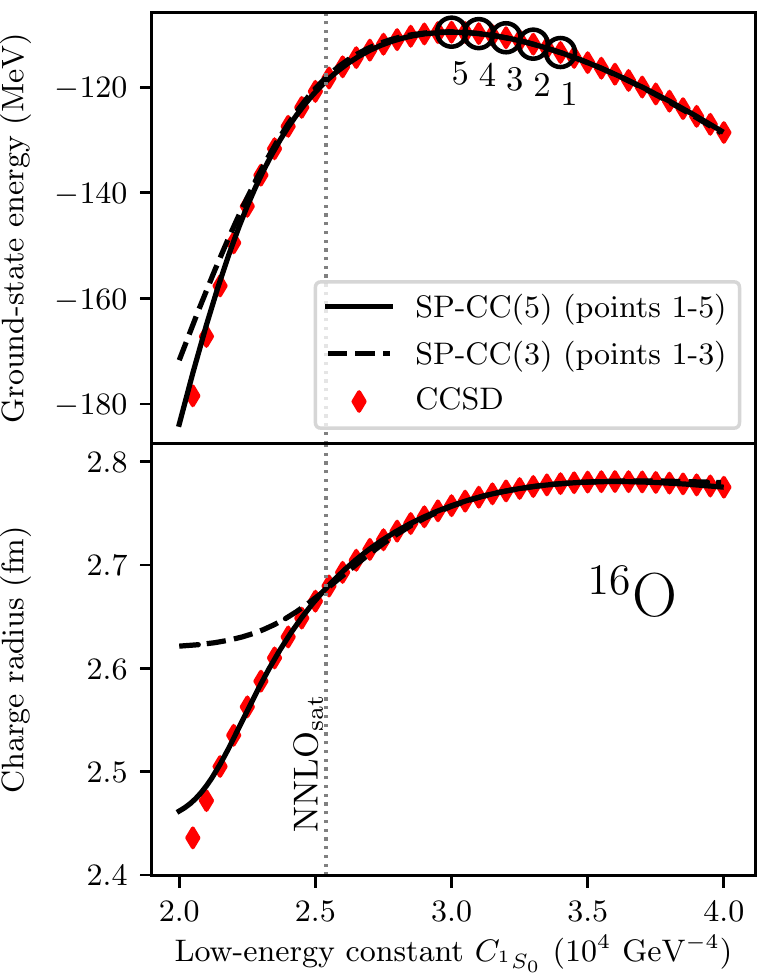}
  \caption{(Color Online) SP-CC results for $^{16}$O, using three or
    five subspace vectors, for different values of the low-energy
    constant (LEC) $C_{^1S_0}$. The red diamonds indicate exact CC
    calculations at the singles and doubles level (CCSD). The
    NNLO$_{\rm sat}$ point is indicated with a dashed vertical line.}
  \label{fig:1S0}
\end{figure}

In this Letter we solve this problem by utilizing eigenvector
continuation~\cite{Frame2018} to develop a subspace-projected
coupled-cluster (SP-CC) method as a fast and accurate approximation to
the corresponding full-space coupled-cluster (CC)
method~\cite{coester1958,coester1960,kuemmel1978,mihaila2000b,dean2004,bartlett2007,hagen2014}. The
SP-CC method generalizes the eigenvector-continuation formalism in
Ref.~\cite{konig2019} to non-Hermitian problems and enables
accelerated computation of nuclear observables across the Segr\'e
chart for any target value $\vec{\alpha}_{\circledcirc}$ of the LECs
in the underlying Hamiltonian. See Fig.~\ref{fig:1S0} for a
demonstration of the SP-CC method applied to $^{16}$O and the
variation of a single LEC (details are given below). We will use SP-CC
to analyze the description of the $^{16}$O ground-state energy and
charge radius across a large domain of relevant LECs. This way we can
for the first time clearly identify the LECs that have the biggest
impact on binding energy and radius predictions, which in turn impacts
saturation properties of nuclear
matter~\cite{ekstrom2017,simonis2017,drischler2019}.

{\it Method.}-- Following Ref.~\cite{konig2019} we start by
representing the chiral Hamiltonian at NNLO $H(\vec{\alpha})$ as a
linear combination with respect to all the LECs $\vec{\alpha}$;
i.e. $H(\vec{\alpha}) = \sum_{i=0}^{N_{\rm LECs}=16} \alpha_i h_i, $
with the zeroth term given by $h_0 = t_{\rm kin} + V_0$ and $\alpha_0
= 1$. Here $t_{\rm kin}$ is the intrinsic kinetic energy and $V_0$
denotes a constant potential term. The analytical form of the NNLO
Hamiltonian is identical to the one of NNLO$_{\rm
  sat}$~\cite{ekstrom2015}, which means that for a particular value
$\vec{\alpha} = \vec{\alpha}_{\star}$ the Hamiltonian
$H(\vec{\alpha}_{\star})$ will reproduce the binding energy and radius
predictions of NNLO$_{\rm sat}$. The SP-CC Hamiltonian for a target
value $\vec{\alpha} = \vec{\alpha}_\circledcirc$ is constructed by
projecting $H(\vec{\alpha}_\circledcirc)$ onto the subspace spanned by
CC wavefunctions obtained at $N_{\rm sub}$ different values for
$\vec{\alpha}$. SP-CC is a controlled approximation to the full-space
CC method, and allows for rapid and accurate solutions to the
many-nucleon problem necessary for statistical computing. In this
Letter we use the CC method in the singles- and doubles approximation
(CCSD).

The workhorse of the CC method is the similarity transformed
Hamiltonian $\overline{H}(\vec{\alpha}) = e^{-T(\vec{\alpha})}
H(\vec{\alpha}) e^{T(\vec{\alpha})}$, where in the CCSD approximation
the cluster operator is truncated at one-particle-one-hole and
two-particle-two-hole excitations, i. e. $T(\vec{\alpha}) =
T_1(\vec{\alpha}) + T_2(\vec{\alpha})$. For clarity, we have indicated
the implicit dependence on $\vec{\alpha}$. The CCSD similarity
transformation is non-unitary and renders $\overline{H}(\vec{\alpha})$
non-Hermitian, and we thus introduce $N_{\rm sub}$ bi-orthogonal left
and right CC ground-states,
\begin{equation} \langle \tilde\Psi \vert = \langle \Phi_0 \vert
(1+\Lambda(\vec{\alpha})) e^{-T(\vec{\alpha})}, \:\: \vert \Psi\rangle =
e^{T(\vec{\alpha})}\vert \Phi_0\rangle.
\label{eq:cc_gs}
\end{equation}
Here $\Lambda(\vec{\alpha}) = \Lambda_1(\vec{\alpha}) +
\Lambda_2(\vec{\alpha})$ is a linear expansion in
one-particle-one-hole and two-particle-two-hole de-excitation
operators, and we have bi-orthonormality according to $\langle
\tilde\Psi \vert\Psi\rangle = 1$.  For notational simplicity we will
from here on omit the explicit $\vec{\alpha}$ dependence in the
(de)-excitation operators and set $T(\vec{\alpha}) = T$ and
$\Lambda(\vec{\alpha}) = \Lambda$, respectively. The reference state
$\vert \Phi_0\rangle$ is built from harmonic oscillator
single-particle states, and we solve the CCSD equations in a
model-space comprising 11 major oscillator shells with a frequency
$\hbar\Omega = 16$~MeV. The matrix-elements of the three-nucleon
interaction that enters the Hamiltonian are truncated by the energy
cut $E_{\rm 3max} \leq 14$.  The CCSD result for $^{16}$O with
NNLO$_{\rm sat}$ in this model-space is $-118.76$~MeV, which is within
1~MeV of the converged CCSD value using a Hartree-Fock basis.

Using the $N_{\rm sub}$ different CCSD ground-state vectors in
Eq.~(\ref{eq:cc_gs}), the matrix elements of the target Hamiltonian in the subspace
and the corresponding norm matrix are given by,
\begin{eqnarray}
\nonumber
\langle {\tilde \Psi}' \vert {H} (\vec{\alpha}_\circledcirc)
  \vert \Psi \rangle & = &  
 \langle \Phi_0 \vert (1+\Lambda')
  e^{-T'} H(\vec{\alpha}_\circledcirc) e^{T}\vert \Phi_0\rangle \\
& = & \langle \Phi_0 \vert (1+\Lambda')
e^{X} \overline{H}(\vec{\alpha}_\circledcirc)\vert \Phi_0\rangle,
\label{eq:h_ij} \\
   \langle\tilde{\Psi}'\vert \Psi\rangle & = & \langle \Phi_0 \vert (1+\Lambda') e^{X} \vert \Phi_0\rangle,
\label{eq:n_ij}
\end{eqnarray}
respectively. Here we also introduced $e^{X} = e^{-T' + T}$, and
$\overline{H}(\vec{\alpha}_\circledcirc)$ is the similarity
transformed target Hamiltonian. The left ground-state $\langle {\tilde
  \Psi}' \vert = \langle \Phi_0 \vert (1+\Lambda') e^{-T'}$ is
obtained from $H(\vec{\alpha}')$, and the right ground-state
$e^{T}\vert \Phi_0\rangle$ is obtained from $H(\vec{\alpha})$,
respectively.  We can now solve the generalized non-Hermitian $N_{\rm
  sub} \times N_{\rm sub}$ eigenvalue problem for the SP-CC target
Hamiltonian to obtain the ground-state energy and wavefunction in the
subspace. With the SP-CC wavefunction we can also calculate the
expectation value of any subspace-projected operator with matrix
elements $\langle {\tilde \Psi}' \vert {O} \vert \Psi \rangle$.
Equations~(\ref{eq:h_ij}) and~(\ref{eq:n_ij}) can be evaluated using
Wick's theorem and closed form algebraic expressions are given in the
Supplementary Material. Note that in general the reference states for
the $N_{\rm sub}$ different subspace CC wavefunctions in
Eq.~(\ref{eq:cc_gs}) are non-orthogonal. This is a non-trivial case
and would require the generalized Wick's
theorem~\cite{hoyos2012,plasser2016} in order to evaluate the matrix
elements of the SP-CC Hamiltonian and the norm matrix.

\begin{figure}[hbt]
  \includegraphics[width=0.9\columnwidth]{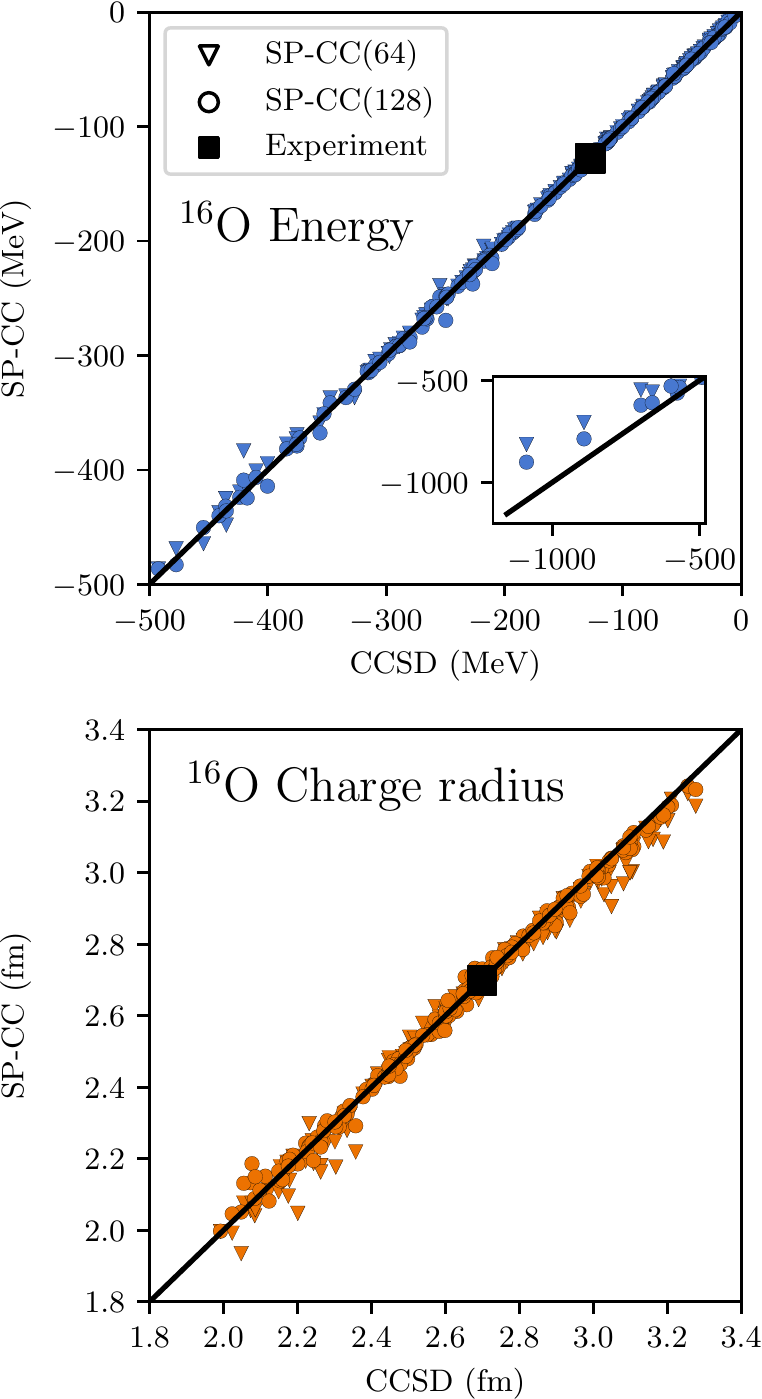}
  \caption{(Color Online) Cross-validation of SP-CC(64) and SP-CC(128)
    for the ground-state energy (top) and charge radius (bottom) using
    200 exact CCSD calculations. The inset shows energy predictions
    below -500 MeV. Only radii for negative-energy states shallower
    than -500 MeV are included.}
  \label{fig:xval}
\end{figure} {\it Results.}-- The SP-CC predictions for the energy and
charge radius in $^{16}$O as a function of the LEC $C_{^1S_0}$ in the
Hamiltonian are shown in Fig~\ref{fig:1S0}. Using $N_{\rm sub}=5$
exact CCSD ground-state vectors, from a small region of $C_{^1S_0}$
values, points 1-5 in Fig.~\ref{fig:1S0}, the SP-CC method
extrapolates to the exact CCSD results across a large $C_{^1S_0}$
range. With $N_{\rm sub}=3$ CCSD vectors, points 1-3 in
Fig.~\ref{fig:1S0}, the radius extrapolation deteriorates far away
from the exact solutions, while the energy predictions remain more
accurate.

We now move to the challenging case where all 16 LECs at NNLO can
vary. In the following we analyze two SP-CC Hamiltonians based on
$N_{\rm sub}=64$ and $N_{\rm sub}=128$ CCSD ground-state vectors,
referred to as SP-CC(64) and SP-CC(128), respectively. The
ground-state vectors are obtained at $N_{\rm sub}$ points in a domain
of LEC values that surrounds the nominal LEC values of NNLO$_{\rm
  sat}$ within $\pm 20\%$ relative variation. This domain spans a
rather large interval of ground-state energies and charge radii in
$^{16}$O. The three-nucleon contact-LEC $c_E \approx 0.0395$ in
NNLO$_{\rm sat}$ is small compared to the values of the remaining 15
LECs, we therefore scaled $c_{E}$ with a factor of 20. In accordance
with observation, we also constrained the leading-order
isospin-breaking $^{1}S_0$ LECs ($\tilde{C}$) to exhibit small
isospin-breaking. We draw $N_{\rm sub}$ values for $\vec{\alpha}$
using a space-filling latin hypercube design and solve for the exact
CCSD wavefunction at each point. We have verified that the SP-CC(64)
and SP-CC(128) Hamiltonians reproduce the energies and radii of the
exact CCSD calculations for all $N_{\rm sub}$ choices of
$\vec{\alpha}$.
\begin{figure*}[hbt]
  \includegraphics[width=0.9\textwidth]{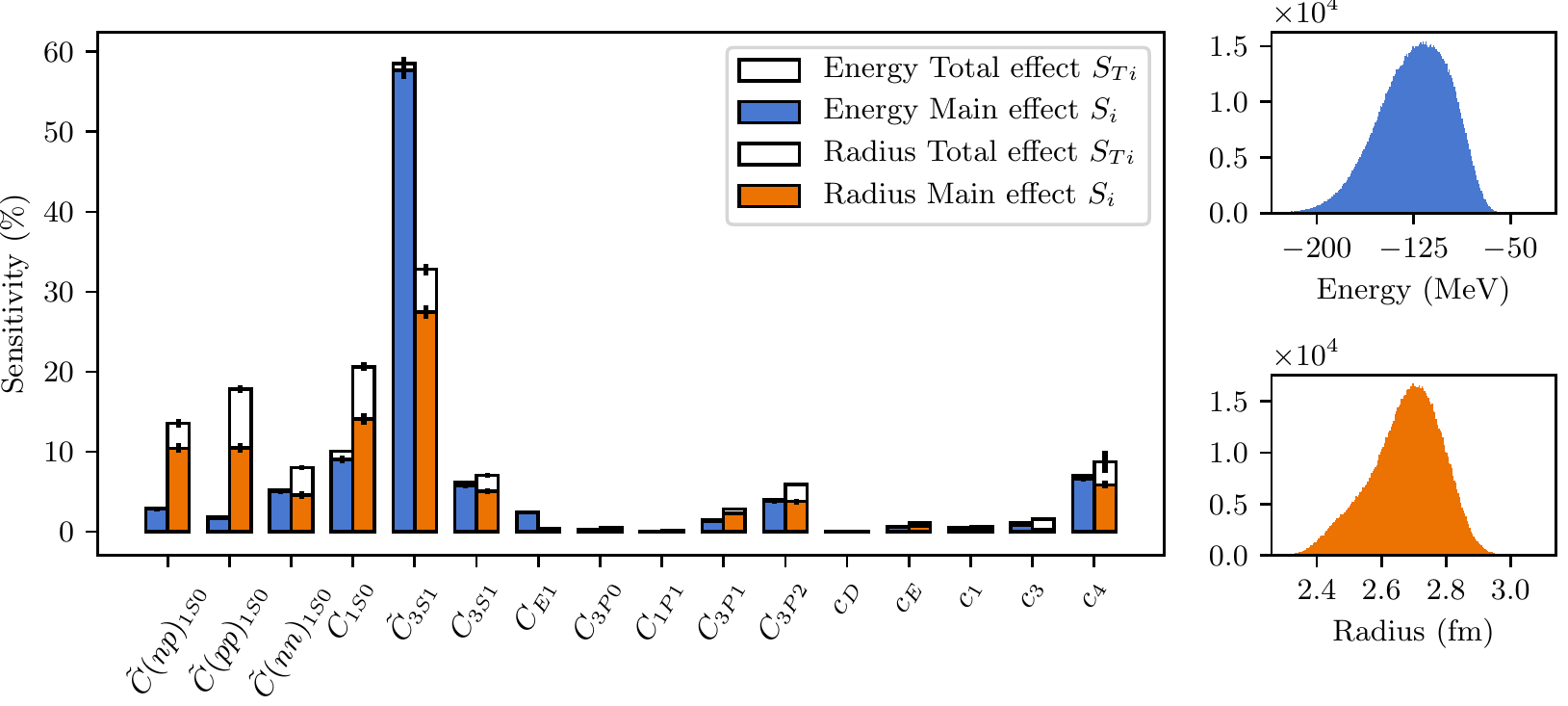}
  \caption{(Color Online) (Left panel) Main and total effects (in \%)
    for the ground-state energy (left bar) and charge radius (right
    bar) in $^{16}$O, grouped per LEC. The main and total effects were
    computed from $(16+1)\cdot2^{16} = 1,114,112$ quasi MC evaluations
    of the SP-CC(64) Hamiltonian. The vertical lines on each bar
    indicate bootstrapped $95\%$ confidence intervals. A larger
    sensitivity value implies that the corresponding LEC is more
    critical for explaining the variance in the model output. (Right
    panels) Histograms of the ground-state energy (top) and
    charge radius (bottom) from which total variances are decomposed.}
  \label{sensitivity}
\end{figure*}

Figure~\ref{fig:xval} shows the cross-validation with respect to an
additional set of 200 randomly drawn exact CCSD calculations in the
same 20\% domain. From the cross-validation we extract a root
mean-squared error (RMSE) of SP-CC(64): 4 MeV and 0.04 fm for the
ground-state energy and charge radius, respectively. With SP-CC(128)
the RMSE values are 3 MeV and 0.02 fm. Using more subspace vectors
gives better predictions. The present results are within the expected
accuracy of CCSD. The non-hermiticity of the CCSD equations yields
SP-CC Hamiltonians that do not obey a variational bound with respect
to the exact CCSD calculations. From Fig.~\ref{fig:xval} we see that
this is a minute effect.

We use SP-CC(64) and global sensitivity analysis (GSA) to analyze how
the \textit{ab initio} predictions for the energy and charge radius in
$^{16}$O explicitly depend on the LECs in the NNLO nuclear
interaction. GSA is a very powerful, although computationally
demanding, method for learning how much each unknown model parameter
contributes to the uncertainty in a model prediction~\cite{Sobol}. As
opposed to an uncertainty analysis, which addresses the question of
how uncertain the prediction itself is.  With SP-CC we can carry out
the large amount of model evaluations that is required to extract
statistically significant GSA results. In the following, we treat the
ground-state energy or radius of $^{16}$O as an output $ Y =
f(\vec{\alpha})$ of a model $f$, given here by the SP-CC(64)
Hamiltonian and its eigendecomposition. In the GSA we decompose the
total variance ${\rm Var}[Y]$ as
\begin{equation}
  {\rm Var}[Y] = \sum_{i=1}^{N_{\rm LECs}} V_i + \sum_{i<j}^{N_{\rm LECs}} V_{ij} + \ldots, 
\end{equation}
where the partial variances are given by
\begin{align}
  \begin{split}
    {}& V_i = {\rm Var}[E_{\vec{\alpha} \sim( \alpha_i)}[Y|\alpha_i]]\\
      {}& V_{ij}= {\rm Var}[E_{\vec{\alpha}\sim (\alpha_i,\alpha_j)}[Y|\alpha_i,\alpha_j]] - V_i - V_j
  \end{split}
\end{align}
and ${\rm Var}[E_{\vec{\alpha}\sim(\alpha_i)}[Y|\alpha_i]]$ denotes
the variance of the conditional expectation of $Y$, and $\vec{\alpha}
\sim (\alpha_i)$ denotes the set of all LECs excluding $\alpha_i$, and
correspondingly for the second-order term. The variance integrals are
evaluated using quasi Monte Carlo (MC) sampling and we extract a 95\%
confidence interval of the final result via bootstrap with 100
re-samples~\cite{Saltelli}. The first- and second-order sensitivity
indices are defined as
  \begin{align}
    \begin{split}
      {}& S_i = \frac{V_i}{{\rm Var}[Y]}, \,\, S_{ij} = \frac{V_{ij}}{{\rm Var}[Y]}.
    \end{split}
  \end{align}

The first-order sensitivity, $S_i$, is often referred to as the main
effect. It apportions the total variance in the model output to an
individual model parameter $\alpha_i$. The higher-order indices,
e.g. $S_{ij}$, apportion the variance in the model output to the
combination of parameters $\alpha_i$ and $\alpha_j$. The number of
higher-order indices grow exponentially with the number of parameters
in the model. Fortunately, it is possible to compute the sum of all
sensitivity indices for each $\alpha_i$, i.e.  $ S_{Ti} = S_i + S_{ij}
+ S_{ijk} + \ldots $. This is referred to as the total effect, and it
quantifies the total sensitivity of ${\rm Var}[Y]$ to parameter
$\alpha_i$ including all of its higher-order parameter
combinations~\cite{Homma}. We always have that $S_{Ti} \geq S_i$, and
equality for purely additive models. In this analysis, we do not
calibrate the model to reproduce data. We study the behavior and
response of the model itself, and assume all LECs to be independent of
each other and uniformly distributed. In future studies one could draw
LECs from a Bayesian posterior distribution.

Figure~\ref{sensitivity} shows the results from the GSA of the
$^{16}$O energy and radius using an SP-CC(64) chiral NNLO
Hamiltonian. To limit the model response of the energy and radius to a
physically reasonable interval we only sample a LEC domain
corresponding to 10\% variation around the NNLO$_{\rm sat}$
values. The LEC $c_E$ is still scaled with a factor of $20$. The MC
sample size required for carrying out a reliable GSA depends on: i)
the complexity of the model, and ii) the number of parameters in the
model. We have to use $(16+1)\cdot 2^{16} = 1,114,112$ quasi MC
samples to extract statistically significant main and total effects of
the energy and radius for all LECs. With SP-CC(64) this took about 1
hour on a standard laptop, while an equivalent set of exact CCSD
computations would require 20 years. We find that $58(1)\%$ of the
variance in the energy can be attributed to the leading order LEC
$\tilde{C}_{^3S_1}$ and all main and total effects are nearly
identical, which signals that the energy is nearly additive in all
LECs. For the radius, the main effects are distributed across several
LECs and they differ from the total effects. Indeed, second-order
correlations between the LECs are responsible for almost 14\% of the
variance in the radius. This reflects part of the challenge of
optimizing chiral NNLO Hamiltonians to reproduce radii of atomic
nuclei and consequently saturation properties of nuclear matter. Our
analysis also reveals that the energy and radius of $^{16}$O are not
sensitive to the short-ranged parts of the three-nucleon interaction
in this domain. Of the long-ranged $\pi N$ LECs, $c_{1,3,4}$, only
$c_4$ exhibits a non-negligible main-effect for the energy and
radius. This LEC contributes to the tensor force. As expected, only
$P-$wave LECs with large spin-weights contribute to the $^{16}$O
wavefunction. There also seems to be a larger sensitivity of the
radius to the isospin- breaking terms in the interaction.  The GSA
results guide future uncertainty reduction efforts for specific
observables by identifying non-influential LECs, which is also useful
for e.g. calibration. The SP-CC method will significantly leverage
statistical computation for analyzing correlations between different
observables in different nuclei across the Segr\'e chart.

{\it Summary and outlook.}-- We have developed the SP-CC method for
evaluating nuclear observables at different values of the LECs in
chiral Hamiltonians at unprecedented speed. With a modest number of
subspace vectors, $N_{\rm sub}=64$, we reached 1\% accuracy relative
to exact CCSD solutions. From a GSA we conclude that the variance of
the ground-state energy in $^{16}$O is additive in all LECs of the
NNLO chiral Hamiltonian, and that the charge radius depends
sensitively on the combination of several LECs. The SP-CC method
enables sophisticated statistical computation~\cite{gelman,
  McDonnell2015, Neufcourt2019, Vernon2010} in \textit{ab initio}
nuclear theory to reveal which new data would best reduce the
uncertainty in Hamiltonian models and for understanding how properties
of atomic nuclei depend on the underlying interaction between protons
and neutrons. The stability of $^{16}$O with respect to breakup into
$^{4}$He clusters is a relevant
example~\cite{carlsson2015,ekstrom2017,contessi2017,bansal2018}. The
SP-CC method also enables straightforward computation of derivatives
with respect to the LECs using e.g. algorithmic differentiation. SP-CC
Hamiltonians occupy very little disk space, and can easily be shared
within the nuclear community.

\begin{acknowledgments}
  We thank Michael Grosskopf, Sebastian K\"onig, Dean Lee, Titus
  Morris, and Thomas Papenbrock for fruitful
  discussions. G.~H. acknowledges the hospitality of Chalmers University
  of Technology where most of this work was carried out. This work was
  supported by the European Research Council (ERC) under the European
  Union’s Horizon 2020 research and innovation programme (Grant
  agreement No. 758027), the Office of Nuclear Physics,
  U.S. Department of Energy, under grants desc0018223 (NUCLEI SciDAC-4
  collaboration) and by the Field Work Proposal ERKBP72 at Oak Ridge
  National Laboratory (ORNL). Computer time was provided by the
  Innovative and Novel Computational Impact on Theory and Experiment
  (INCITE) program. This research used resources of the Oak Ridge
  Leadership Computing Facility located at ORNL, which is supported by
  the Office of Science of the Department of Energy under Contract
  No. DE-AC05-00OR22725.
\end{acknowledgments}
\bibliography{refs}

\begin{thebibliography}{48}%
\makeatletter
\providecommand \@ifxundefined [1]{%
 \@ifx{#1\undefined}
}%
\providecommand \@ifnum [1]{%
 \ifnum #1\expandafter \@firstoftwo
 \else \expandafter \@secondoftwo
 \fi
}%
\providecommand \@ifx [1]{%
 \ifx #1\expandafter \@firstoftwo
 \else \expandafter \@secondoftwo
 \fi
}%
\providecommand \natexlab [1]{#1}%
\providecommand \enquote  [1]{``#1''}%
\providecommand \bibnamefont  [1]{#1}%
\providecommand \bibfnamefont [1]{#1}%
\providecommand \citenamefont [1]{#1}%
\providecommand \href@noop [0]{\@secondoftwo}%
\providecommand \href [0]{\begingroup \@sanitize@url \@href}%
\providecommand \@href[1]{\@@startlink{#1}\@@href}%
\providecommand \@@href[1]{\endgroup#1\@@endlink}%
\providecommand \@sanitize@url [0]{\catcode `\\12\catcode `\$12\catcode
  `\&12\catcode `\#12\catcode `\^12\catcode `\_12\catcode `\%12\relax}%
\providecommand \@@startlink[1]{}%
\providecommand \@@endlink[0]{}%
\providecommand \url  [0]{\begingroup\@sanitize@url \@url }%
\providecommand \@url [1]{\endgroup\@href {#1}{\urlprefix }}%
\providecommand \urlprefix  [0]{URL }%
\providecommand \Eprint [0]{\href }%
\providecommand \doibase [0]{http://dx.doi.org/}%
\providecommand \selectlanguage [0]{\@gobble}%
\providecommand \bibinfo  [0]{\@secondoftwo}%
\providecommand \bibfield  [0]{\@secondoftwo}%
\providecommand \translation [1]{[#1]}%
\providecommand \BibitemOpen [0]{}%
\providecommand \bibitemStop [0]{}%
\providecommand \bibitemNoStop [0]{.\EOS\space}%
\providecommand \EOS [0]{\spacefactor3000\relax}%
\providecommand \BibitemShut  [1]{\csname bibitem#1\endcsname}%
\let\auto@bib@innerbib\@empty
\bibitem [{\citenamefont {Hergert}\ \emph {et~al.}(2014)\citenamefont
  {Hergert}, \citenamefont {Bogner}, \citenamefont {Morris}, \citenamefont
  {Binder}, \citenamefont {Calci}, \citenamefont {Langhammer},\ and\
  \citenamefont {Roth}}]{hergert2014}%
  \BibitemOpen
  \bibfield  {author} {\bibinfo {author} {\bibfnamefont {H.}~\bibnamefont
  {Hergert}}, \bibinfo {author} {\bibfnamefont {S.~K.}\ \bibnamefont {Bogner}},
  \bibinfo {author} {\bibfnamefont {T.~D.}\ \bibnamefont {Morris}}, \bibinfo
  {author} {\bibfnamefont {S.}~\bibnamefont {Binder}}, \bibinfo {author}
  {\bibfnamefont {A.}~\bibnamefont {Calci}}, \bibinfo {author} {\bibfnamefont
  {J.}~\bibnamefont {Langhammer}}, \ and\ \bibinfo {author} {\bibfnamefont
  {R.}~\bibnamefont {Roth}},\ }\bibfield  {title} {\enquote {\bibinfo {title}
  {\textit{Ab initio} multireference in-medium similarity renormalization group
  calculations of even calcium and nickel isotopes},}\ }\href {\doibase
  10.1103/PhysRevC.90.041302} {\bibfield  {journal} {\bibinfo  {journal} {Phys.
  Rev. C}\ }\textbf {\bibinfo {volume} {90}},\ \bibinfo {pages} {041302}
  (\bibinfo {year} {2014})}\BibitemShut {NoStop}%
\bibitem [{\citenamefont {Elhatisari}\ \emph {et~al.}(2015)\citenamefont
  {Elhatisari}, \citenamefont {Lee}, \citenamefont {Rupak}, \citenamefont
  {Epelbaum}, \citenamefont {Krebs}, \citenamefont {L{\"a}hde}, \citenamefont
  {Luu},\ and\ \citenamefont {Mei{\ss}ner}}]{elhatisari2015}%
  \BibitemOpen
  \bibfield  {author} {\bibinfo {author} {\bibfnamefont {S.}~\bibnamefont
  {Elhatisari}}, \bibinfo {author} {\bibfnamefont {D.}~\bibnamefont {Lee}},
  \bibinfo {author} {\bibfnamefont {G.}~\bibnamefont {Rupak}}, \bibinfo
  {author} {\bibfnamefont {E.}~\bibnamefont {Epelbaum}}, \bibinfo {author}
  {\bibfnamefont {H.}~\bibnamefont {Krebs}}, \bibinfo {author} {\bibfnamefont
  {T.~A.}\ \bibnamefont {L{\"a}hde}}, \bibinfo {author} {\bibfnamefont
  {T.}~\bibnamefont {Luu}}, \ and\ \bibinfo {author} {\bibfnamefont {U.-G.}\
  \bibnamefont {Mei{\ss}ner}},\ }\bibfield  {title} {\enquote {\bibinfo {title}
  {Ab initio alpha--alpha scattering},}\ }\href {\doibase 10.1038/nature16067}
  {\bibfield  {journal} {\bibinfo  {journal} {Nature}\ }\textbf {\bibinfo
  {volume} {528}},\ \bibinfo {pages} {111--114} (\bibinfo {year}
  {2015})}\BibitemShut {NoStop}%
\bibitem [{\citenamefont {Rosenbusch}\ \emph {et~al.}(2015)\citenamefont
  {Rosenbusch}, \citenamefont {Ascher}, \citenamefont {Atanasov}, \citenamefont
  {Barbieri}, \citenamefont {Beck}, \citenamefont {Blaum}, \citenamefont
  {Borgmann}, \citenamefont {Breitenfeldt}, \citenamefont {Cakirli},
  \citenamefont {Cipollone}, \citenamefont {George}, \citenamefont {Herfurth},
  \citenamefont {Kowalska}, \citenamefont {Kreim}, \citenamefont {Lunney},
  \citenamefont {Manea}, \citenamefont {Navr\'atil}, \citenamefont {Neidherr},
  \citenamefont {Schweikhard}, \citenamefont {Som\`a}, \citenamefont {Stanja},
  \citenamefont {Wienholtz}, \citenamefont {Wolf},\ and\ \citenamefont
  {Zuber}}]{rosenbusch2015}%
  \BibitemOpen
  \bibfield  {author} {\bibinfo {author} {\bibfnamefont {M.}~\bibnamefont
  {Rosenbusch}}, \bibinfo {author} {\bibfnamefont {P.}~\bibnamefont {Ascher}},
  \bibinfo {author} {\bibfnamefont {D.}~\bibnamefont {Atanasov}}, \bibinfo
  {author} {\bibfnamefont {C.}~\bibnamefont {Barbieri}}, \bibinfo {author}
  {\bibfnamefont {D.}~\bibnamefont {Beck}}, \bibinfo {author} {\bibfnamefont
  {K.}~\bibnamefont {Blaum}}, \bibinfo {author} {\bibfnamefont {Ch.}\
  \bibnamefont {Borgmann}}, \bibinfo {author} {\bibfnamefont {M.}~\bibnamefont
  {Breitenfeldt}}, \bibinfo {author} {\bibfnamefont {R.~B.}\ \bibnamefont
  {Cakirli}}, \bibinfo {author} {\bibfnamefont {A.}~\bibnamefont {Cipollone}},
  \bibinfo {author} {\bibfnamefont {S.}~\bibnamefont {George}}, \bibinfo
  {author} {\bibfnamefont {F.}~\bibnamefont {Herfurth}}, \bibinfo {author}
  {\bibfnamefont {M.}~\bibnamefont {Kowalska}}, \bibinfo {author}
  {\bibfnamefont {S.}~\bibnamefont {Kreim}}, \bibinfo {author} {\bibfnamefont
  {D.}~\bibnamefont {Lunney}}, \bibinfo {author} {\bibfnamefont
  {V.}~\bibnamefont {Manea}}, \bibinfo {author} {\bibfnamefont
  {P.}~\bibnamefont {Navr\'atil}}, \bibinfo {author} {\bibfnamefont
  {D.}~\bibnamefont {Neidherr}}, \bibinfo {author} {\bibfnamefont
  {L.}~\bibnamefont {Schweikhard}}, \bibinfo {author} {\bibfnamefont
  {V.}~\bibnamefont {Som\`a}}, \bibinfo {author} {\bibfnamefont
  {J.}~\bibnamefont {Stanja}}, \bibinfo {author} {\bibfnamefont
  {F.}~\bibnamefont {Wienholtz}}, \bibinfo {author} {\bibfnamefont {R.~N.}\
  \bibnamefont {Wolf}}, \ and\ \bibinfo {author} {\bibfnamefont
  {K.}~\bibnamefont {Zuber}},\ }\bibfield  {title} {\enquote {\bibinfo {title}
  {Probing the $n=32$ shell closure below the magic proton number $z=20$: Mass
  measurements of the exotic isotopes $^{52,53}\mathrm{K}$},}\ }\href {\doibase
  10.1103/PhysRevLett.114.202501} {\bibfield  {journal} {\bibinfo  {journal}
  {Phys. Rev. Lett.}\ }\textbf {\bibinfo {volume} {114}},\ \bibinfo {pages}
  {202501} (\bibinfo {year} {2015})}\BibitemShut {NoStop}%
\bibitem [{\citenamefont {{Hagen}}\ \emph {et~al.}(2016)\citenamefont
  {{Hagen}}, \citenamefont {{Ekstr{\"o}m}}, \citenamefont {{Forss{\'e}n}},
  \citenamefont {{Jansen}}, \citenamefont {{Nazarewicz}}, \citenamefont
  {{Papenbrock}}, \citenamefont {{Wendt}}, \citenamefont {{Bacca}},
  \citenamefont {{Barnea}}, \citenamefont {{Carlsson}}, \citenamefont
  {{Drischler}}, \citenamefont {{Hebeler}}, \citenamefont {{Hjorth-Jensen}},
  \citenamefont {{Miorelli}}, \citenamefont {{Orlandini}}, \citenamefont
  {{Schwenk}},\ and\ \citenamefont {{Simonis}}}]{hagen2015}%
  \BibitemOpen
  \bibfield  {author} {\bibinfo {author} {\bibfnamefont {G.}~\bibnamefont
  {{Hagen}}}, \bibinfo {author} {\bibfnamefont {A.}~\bibnamefont
  {{Ekstr{\"o}m}}}, \bibinfo {author} {\bibfnamefont {C.}~\bibnamefont
  {{Forss{\'e}n}}}, \bibinfo {author} {\bibfnamefont {G.~R.}\ \bibnamefont
  {{Jansen}}}, \bibinfo {author} {\bibfnamefont {W.}~\bibnamefont
  {{Nazarewicz}}}, \bibinfo {author} {\bibfnamefont {T.}~\bibnamefont
  {{Papenbrock}}}, \bibinfo {author} {\bibfnamefont {K.~A.}\ \bibnamefont
  {{Wendt}}}, \bibinfo {author} {\bibfnamefont {S.}~\bibnamefont {{Bacca}}},
  \bibinfo {author} {\bibfnamefont {N.}~\bibnamefont {{Barnea}}}, \bibinfo
  {author} {\bibfnamefont {B.}~\bibnamefont {{Carlsson}}}, \bibinfo {author}
  {\bibfnamefont {C.}~\bibnamefont {{Drischler}}}, \bibinfo {author}
  {\bibfnamefont {K.}~\bibnamefont {{Hebeler}}}, \bibinfo {author}
  {\bibfnamefont {M.}~\bibnamefont {{Hjorth-Jensen}}}, \bibinfo {author}
  {\bibfnamefont {M.}~\bibnamefont {{Miorelli}}}, \bibinfo {author}
  {\bibfnamefont {G.}~\bibnamefont {{Orlandini}}}, \bibinfo {author}
  {\bibfnamefont {A.}~\bibnamefont {{Schwenk}}}, \ and\ \bibinfo {author}
  {\bibfnamefont {J.}~\bibnamefont {{Simonis}}},\ }\bibfield  {title} {\enquote
  {\bibinfo {title} {{Neutron and weak-charge distributions of the $^{48}$Ca
  nucleus}},}\ }\href {\doibase 10.1038/nphys3529} {\bibfield  {journal}
  {\bibinfo  {journal} {Nature Phys.}\ }\textbf {\bibinfo {volume} {12}},\
  \bibinfo {pages} {186} (\bibinfo {year} {2016})}\BibitemShut {NoStop}%
\bibitem [{\citenamefont {{Garcia Ruiz}}\ \emph {et~al.}(2016)\citenamefont
  {{Garcia Ruiz}}, \citenamefont {{Bissell}}, \citenamefont {{Blaum}},
  \citenamefont {{Ekstr\"om}}, \citenamefont {{Fr\"ommgen}}, \citenamefont
  {{Hagen}}, \citenamefont {{Hammen}}, \citenamefont {{Hebeler}}, \citenamefont
  {{Holt}}, \citenamefont {{Jansen}}, \citenamefont {{Kowalska}}, \citenamefont
  {{Kreim}}, \citenamefont {{Nazarewicz}}, \citenamefont {{Neugart}},
  \citenamefont {{Neyens}}, \citenamefont {{N\"ortersh\"auser}}, \citenamefont
  {{Papenbrock}}, \citenamefont {{Papuga}}, \citenamefont {{Schwenk}},
  \citenamefont {{Simonis}}, \citenamefont {{Wendt}},\ and\ \citenamefont
  {{Yordanov}}}]{garciaruiz2016}%
  \BibitemOpen
  \bibfield  {author} {\bibinfo {author} {\bibfnamefont {R.~F.}\ \bibnamefont
  {{Garcia Ruiz}}}, \bibinfo {author} {\bibfnamefont {M.~L.}\ \bibnamefont
  {{Bissell}}}, \bibinfo {author} {\bibfnamefont {K.}~\bibnamefont {{Blaum}}},
  \bibinfo {author} {\bibfnamefont {A.}~\bibnamefont {{Ekstr\"om}}}, \bibinfo
  {author} {\bibfnamefont {N.}~\bibnamefont {{Fr\"ommgen}}}, \bibinfo {author}
  {\bibfnamefont {G.}~\bibnamefont {{Hagen}}}, \bibinfo {author} {\bibfnamefont
  {M.}~\bibnamefont {{Hammen}}}, \bibinfo {author} {\bibfnamefont
  {K.}~\bibnamefont {{Hebeler}}}, \bibinfo {author} {\bibfnamefont {J.~D.}\
  \bibnamefont {{Holt}}}, \bibinfo {author} {\bibfnamefont {G.~R.}\
  \bibnamefont {{Jansen}}}, \bibinfo {author} {\bibfnamefont {M.}~\bibnamefont
  {{Kowalska}}}, \bibinfo {author} {\bibfnamefont {K.}~\bibnamefont {{Kreim}}},
  \bibinfo {author} {\bibfnamefont {W.}~\bibnamefont {{Nazarewicz}}}, \bibinfo
  {author} {\bibfnamefont {R.}~\bibnamefont {{Neugart}}}, \bibinfo {author}
  {\bibfnamefont {G.}~\bibnamefont {{Neyens}}}, \bibinfo {author}
  {\bibfnamefont {W.}~\bibnamefont {{N\"ortersh\"auser}}}, \bibinfo {author}
  {\bibfnamefont {T.}~\bibnamefont {{Papenbrock}}}, \bibinfo {author}
  {\bibfnamefont {J.}~\bibnamefont {{Papuga}}}, \bibinfo {author}
  {\bibfnamefont {A.}~\bibnamefont {{Schwenk}}}, \bibinfo {author}
  {\bibfnamefont {J.}~\bibnamefont {{Simonis}}}, \bibinfo {author}
  {\bibfnamefont {K.~A.}\ \bibnamefont {{Wendt}}}, \ and\ \bibinfo {author}
  {\bibfnamefont {D.~T.}\ \bibnamefont {{Yordanov}}},\ }\bibfield  {title}
  {\enquote {\bibinfo {title} {Unexpectedly large charge radii of neutron-rich
  calcium isotopes},}\ }\href {\doibase 10.1038/nphys3645} {\bibfield
  {journal} {\bibinfo  {journal} {Nature Physics}\ } (\bibinfo {year} {2016}),\
  10.1038/nphys3645},\ \Eprint {http://arxiv.org/abs/1602.07906} {1602.07906}
  \BibitemShut {NoStop}%
\bibitem [{\citenamefont {Lapoux}\ \emph {et~al.}(2016)\citenamefont {Lapoux},
  \citenamefont {Som\`a}, \citenamefont {Barbieri}, \citenamefont {Hergert},
  \citenamefont {Holt},\ and\ \citenamefont {Stroberg}}]{lapoux2016}%
  \BibitemOpen
  \bibfield  {author} {\bibinfo {author} {\bibfnamefont {V.}~\bibnamefont
  {Lapoux}}, \bibinfo {author} {\bibfnamefont {V.}~\bibnamefont {Som\`a}},
  \bibinfo {author} {\bibfnamefont {C.}~\bibnamefont {Barbieri}}, \bibinfo
  {author} {\bibfnamefont {H.}~\bibnamefont {Hergert}}, \bibinfo {author}
  {\bibfnamefont {J.~D.}\ \bibnamefont {Holt}}, \ and\ \bibinfo {author}
  {\bibfnamefont {S.~R.}\ \bibnamefont {Stroberg}},\ }\bibfield  {title}
  {\enquote {\bibinfo {title} {Radii and binding energies in oxygen isotopes: A
  challenge for nuclear forces},}\ }\href {\doibase
  10.1103/PhysRevLett.117.052501} {\bibfield  {journal} {\bibinfo  {journal}
  {Phys. Rev. Lett.}\ }\textbf {\bibinfo {volume} {117}},\ \bibinfo {pages}
  {052501} (\bibinfo {year} {2016})}\BibitemShut {NoStop}%
\bibitem [{\citenamefont {Hagen}\ \emph {et~al.}(2016)\citenamefont {Hagen},
  \citenamefont {Jansen},\ and\ \citenamefont {Papenbrock}}]{hagen2016b}%
  \BibitemOpen
  \bibfield  {author} {\bibinfo {author} {\bibfnamefont {G.}~\bibnamefont
  {Hagen}}, \bibinfo {author} {\bibfnamefont {G.~R.}\ \bibnamefont {Jansen}}, \
  and\ \bibinfo {author} {\bibfnamefont {T.}~\bibnamefont {Papenbrock}},\
  }\bibfield  {title} {\enquote {\bibinfo {title} {Structure of
  $^{78}\mathrm{Ni}$ from first-principles computations},}\ }\href {\doibase
  10.1103/PhysRevLett.117.172501} {\bibfield  {journal} {\bibinfo  {journal}
  {Phys. Rev. Lett.}\ }\textbf {\bibinfo {volume} {117}},\ \bibinfo {pages}
  {172501} (\bibinfo {year} {2016})}\BibitemShut {NoStop}%
\bibitem [{\citenamefont {Simonis}\ \emph {et~al.}(2017)\citenamefont
  {Simonis}, \citenamefont {Stroberg}, \citenamefont {Hebeler}, \citenamefont
  {Holt},\ and\ \citenamefont {Schwenk}}]{simonis2017}%
  \BibitemOpen
  \bibfield  {author} {\bibinfo {author} {\bibfnamefont {J.}~\bibnamefont
  {Simonis}}, \bibinfo {author} {\bibfnamefont {S.~R.}\ \bibnamefont
  {Stroberg}}, \bibinfo {author} {\bibfnamefont {K.}~\bibnamefont {Hebeler}},
  \bibinfo {author} {\bibfnamefont {J.~D.}\ \bibnamefont {Holt}}, \ and\
  \bibinfo {author} {\bibfnamefont {A.}~\bibnamefont {Schwenk}},\ }\bibfield
  {title} {\enquote {\bibinfo {title} {Saturation with chiral interactions and
  consequences for finite nuclei},}\ }\href {\doibase
  10.1103/PhysRevC.96.014303} {\bibfield  {journal} {\bibinfo  {journal} {Phys.
  Rev. C}\ }\textbf {\bibinfo {volume} {96}},\ \bibinfo {pages} {014303}
  (\bibinfo {year} {2017})}\BibitemShut {NoStop}%
\bibitem [{\citenamefont {Duguet}\ \emph {et~al.}(2017)\citenamefont {Duguet},
  \citenamefont {Som\`a}, \citenamefont {Lecluse}, \citenamefont {Barbieri},\
  and\ \citenamefont {Navr\'atil}}]{duguet2017}%
  \BibitemOpen
  \bibfield  {author} {\bibinfo {author} {\bibfnamefont {T.}~\bibnamefont
  {Duguet}}, \bibinfo {author} {\bibfnamefont {V.}~\bibnamefont {Som\`a}},
  \bibinfo {author} {\bibfnamefont {S.}~\bibnamefont {Lecluse}}, \bibinfo
  {author} {\bibfnamefont {C.}~\bibnamefont {Barbieri}}, \ and\ \bibinfo
  {author} {\bibfnamefont {P.}~\bibnamefont {Navr\'atil}},\ }\bibfield  {title}
  {\enquote {\bibinfo {title} {{Ab initio calculation of the potential bubble
  nucleus $^{34}$Si}},}\ }\href {\doibase 10.1103/PhysRevC.95.034319}
  {\bibfield  {journal} {\bibinfo  {journal} {Phys. Rev. C}\ }\textbf {\bibinfo
  {volume} {95}},\ \bibinfo {pages} {034319} (\bibinfo {year}
  {2017})}\BibitemShut {NoStop}%
\bibitem [{\citenamefont {{Morris}}\ \emph {et~al.}(2018)\citenamefont
  {{Morris}}, \citenamefont {{Simonis}}, \citenamefont {{Stroberg}},
  \citenamefont {{Stumpf}}, \citenamefont {{Hagen}}, \citenamefont {{Holt}},
  \citenamefont {{Jansen}}, \citenamefont {{Papenbrock}}, \citenamefont
  {{Roth}},\ and\ \citenamefont {{Schwenk}}}]{morris2017}%
  \BibitemOpen
  \bibfield  {author} {\bibinfo {author} {\bibfnamefont {T.~D.}\ \bibnamefont
  {{Morris}}}, \bibinfo {author} {\bibfnamefont {J.}~\bibnamefont {{Simonis}}},
  \bibinfo {author} {\bibfnamefont {S.~R.}\ \bibnamefont {{Stroberg}}},
  \bibinfo {author} {\bibfnamefont {C.}~\bibnamefont {{Stumpf}}}, \bibinfo
  {author} {\bibfnamefont {G.}~\bibnamefont {{Hagen}}}, \bibinfo {author}
  {\bibfnamefont {J.~D.}\ \bibnamefont {{Holt}}}, \bibinfo {author}
  {\bibfnamefont {G.~R.}\ \bibnamefont {{Jansen}}}, \bibinfo {author}
  {\bibfnamefont {T.}~\bibnamefont {{Papenbrock}}}, \bibinfo {author}
  {\bibfnamefont {R.}~\bibnamefont {{Roth}}}, \ and\ \bibinfo {author}
  {\bibfnamefont {A.}~\bibnamefont {{Schwenk}}},\ }\bibfield  {title} {\enquote
  {\bibinfo {title} {Structure of the lightest tin isotopes},}\ }\href
  {\doibase 10.1103/PhysRevLett.120.152503} {\bibfield  {journal} {\bibinfo
  {journal} {Phys. Rev. Lett.}\ }\textbf {\bibinfo {volume} {120}},\ \bibinfo
  {pages} {152503} (\bibinfo {year} {2018})}\BibitemShut {NoStop}%
\bibitem [{\citenamefont {Lu}\ \emph {et~al.}(2019)\citenamefont {Lu},
  \citenamefont {Li}, \citenamefont {Elhatisari}, \citenamefont {Lee},
  \citenamefont {Epelbaum},\ and\ \citenamefont {Meißner}}]{lu2019}%
  \BibitemOpen
  \bibfield  {author} {\bibinfo {author} {\bibfnamefont {Bing-Nan}\
  \bibnamefont {Lu}}, \bibinfo {author} {\bibfnamefont {Ning}\ \bibnamefont
  {Li}}, \bibinfo {author} {\bibfnamefont {Serdar}\ \bibnamefont {Elhatisari}},
  \bibinfo {author} {\bibfnamefont {Dean}\ \bibnamefont {Lee}}, \bibinfo
  {author} {\bibfnamefont {Evgeny}\ \bibnamefont {Epelbaum}}, \ and\ \bibinfo
  {author} {\bibfnamefont {Ulf-G.}\ \bibnamefont {Meißner}},\ }\bibfield
  {title} {\enquote {\bibinfo {title} {Essential elements for nuclear
  binding},}\ }\href {\doibase https://doi.org/10.1016/j.physletb.2019.134863}
  {\bibfield  {journal} {\bibinfo  {journal} {Physics Letters B}\ }\textbf
  {\bibinfo {volume} {797}},\ \bibinfo {pages} {134863} (\bibinfo {year}
  {2019})}\BibitemShut {NoStop}%
\bibitem [{\citenamefont {Liu}\ \emph {et~al.}(2019)\citenamefont {Liu},
  \citenamefont {Obertelli}, \citenamefont {Doornenbal}, \citenamefont
  {Bertulani}, \citenamefont {Hagen}, \citenamefont {Holt}, \citenamefont
  {Jansen}, \citenamefont {Morris}, \citenamefont {Schwenk}, \citenamefont
  {Stroberg}, \citenamefont {Achouri}, \citenamefont {Baba}, \citenamefont
  {Browne}, \citenamefont {Calvet}, \citenamefont {Ch\^ateau}, \citenamefont
  {Chen}, \citenamefont {Chiga}, \citenamefont {Corsi}, \citenamefont
  {Cort\'es}, \citenamefont {Delbart}, \citenamefont {Gheller}, \citenamefont
  {Giganon}, \citenamefont {Gillibert}, \citenamefont {Hilaire}, \citenamefont
  {Isobe}, \citenamefont {Kobayashi}, \citenamefont {Kubota}, \citenamefont
  {Lapoux}, \citenamefont {Motobayashi}, \citenamefont {Murray}, \citenamefont
  {Otsu}, \citenamefont {Panin}, \citenamefont {Paul}, \citenamefont
  {Rodriguez}, \citenamefont {Sakurai}, \citenamefont {Sasano}, \citenamefont
  {Steppenbeck}, \citenamefont {Stuhl}, \citenamefont {Sun}, \citenamefont
  {Togano}, \citenamefont {Uesaka}, \citenamefont {Wimmer}, \citenamefont
  {Yoneda}, \citenamefont {Aktas}, \citenamefont {Aumann}, \citenamefont
  {Chung}, \citenamefont {Flavigny}, \citenamefont {Franchoo}, \citenamefont
  {Ga\ifmmode \check{s}\else \v{s}\fi{}pari\ifmmode~\acute{c}\else \'{c}\fi{}},
  \citenamefont {Gerst}, \citenamefont {Gibelin}, \citenamefont {Hahn},
  \citenamefont {Kim}, \citenamefont {Koiwai}, \citenamefont {Kondo},
  \citenamefont {Koseoglou}, \citenamefont {Lee}, \citenamefont {Lehr},
  \citenamefont {Linh}, \citenamefont {Lokotko}, \citenamefont {MacCormick},
  \citenamefont {Moschner}, \citenamefont {Nakamura}, \citenamefont {Park},
  \citenamefont {Rossi}, \citenamefont {Sahin}, \citenamefont {Sohler},
  \citenamefont {S\"oderstr\"om}, \citenamefont {Takeuchi}, \citenamefont
  {T\"ornqvist}, \citenamefont {Vaquero}, \citenamefont {Wagner}, \citenamefont
  {Wang}, \citenamefont {Werner}, \citenamefont {Xu}, \citenamefont {Yamada},
  \citenamefont {Yan}, \citenamefont {Yang}, \citenamefont {Yasuda},\ and\
  \citenamefont {Zanetti}}]{liu2019}%
  \BibitemOpen
  \bibfield  {author} {\bibinfo {author} {\bibfnamefont {H.~N.}\ \bibnamefont
  {Liu}}, \bibinfo {author} {\bibfnamefont {A.}~\bibnamefont {Obertelli}},
  \bibinfo {author} {\bibfnamefont {P.}~\bibnamefont {Doornenbal}}, \bibinfo
  {author} {\bibfnamefont {C.~A.}\ \bibnamefont {Bertulani}}, \bibinfo {author}
  {\bibfnamefont {G.}~\bibnamefont {Hagen}}, \bibinfo {author} {\bibfnamefont
  {J.~D.}\ \bibnamefont {Holt}}, \bibinfo {author} {\bibfnamefont {G.~R.}\
  \bibnamefont {Jansen}}, \bibinfo {author} {\bibfnamefont {T.~D.}\
  \bibnamefont {Morris}}, \bibinfo {author} {\bibfnamefont {A.}~\bibnamefont
  {Schwenk}}, \bibinfo {author} {\bibfnamefont {R.}~\bibnamefont {Stroberg}},
  \bibinfo {author} {\bibfnamefont {N.}~\bibnamefont {Achouri}}, \bibinfo
  {author} {\bibfnamefont {H.}~\bibnamefont {Baba}}, \bibinfo {author}
  {\bibfnamefont {F.}~\bibnamefont {Browne}}, \bibinfo {author} {\bibfnamefont
  {D.}~\bibnamefont {Calvet}}, \bibinfo {author} {\bibfnamefont
  {F.}~\bibnamefont {Ch\^ateau}}, \bibinfo {author} {\bibfnamefont
  {S.}~\bibnamefont {Chen}}, \bibinfo {author} {\bibfnamefont {N.}~\bibnamefont
  {Chiga}}, \bibinfo {author} {\bibfnamefont {A.}~\bibnamefont {Corsi}},
  \bibinfo {author} {\bibfnamefont {M.~L.}\ \bibnamefont {Cort\'es}}, \bibinfo
  {author} {\bibfnamefont {A.}~\bibnamefont {Delbart}}, \bibinfo {author}
  {\bibfnamefont {J.-M.}\ \bibnamefont {Gheller}}, \bibinfo {author}
  {\bibfnamefont {A.}~\bibnamefont {Giganon}}, \bibinfo {author} {\bibfnamefont
  {A.}~\bibnamefont {Gillibert}}, \bibinfo {author} {\bibfnamefont
  {C.}~\bibnamefont {Hilaire}}, \bibinfo {author} {\bibfnamefont
  {T.}~\bibnamefont {Isobe}}, \bibinfo {author} {\bibfnamefont
  {T.}~\bibnamefont {Kobayashi}}, \bibinfo {author} {\bibfnamefont
  {Y.}~\bibnamefont {Kubota}}, \bibinfo {author} {\bibfnamefont
  {V.}~\bibnamefont {Lapoux}}, \bibinfo {author} {\bibfnamefont
  {T.}~\bibnamefont {Motobayashi}}, \bibinfo {author} {\bibfnamefont
  {I.}~\bibnamefont {Murray}}, \bibinfo {author} {\bibfnamefont
  {H.}~\bibnamefont {Otsu}}, \bibinfo {author} {\bibfnamefont {V.}~\bibnamefont
  {Panin}}, \bibinfo {author} {\bibfnamefont {N.}~\bibnamefont {Paul}},
  \bibinfo {author} {\bibfnamefont {W.}~\bibnamefont {Rodriguez}}, \bibinfo
  {author} {\bibfnamefont {H.}~\bibnamefont {Sakurai}}, \bibinfo {author}
  {\bibfnamefont {M.}~\bibnamefont {Sasano}}, \bibinfo {author} {\bibfnamefont
  {D.}~\bibnamefont {Steppenbeck}}, \bibinfo {author} {\bibfnamefont
  {L.}~\bibnamefont {Stuhl}}, \bibinfo {author} {\bibfnamefont {Y.~L.}\
  \bibnamefont {Sun}}, \bibinfo {author} {\bibfnamefont {Y.}~\bibnamefont
  {Togano}}, \bibinfo {author} {\bibfnamefont {T.}~\bibnamefont {Uesaka}},
  \bibinfo {author} {\bibfnamefont {K.}~\bibnamefont {Wimmer}}, \bibinfo
  {author} {\bibfnamefont {K.}~\bibnamefont {Yoneda}}, \bibinfo {author}
  {\bibfnamefont {O.}~\bibnamefont {Aktas}}, \bibinfo {author} {\bibfnamefont
  {T.}~\bibnamefont {Aumann}}, \bibinfo {author} {\bibfnamefont {L.~X.}\
  \bibnamefont {Chung}}, \bibinfo {author} {\bibfnamefont {F.}~\bibnamefont
  {Flavigny}}, \bibinfo {author} {\bibfnamefont {S.}~\bibnamefont {Franchoo}},
  \bibinfo {author} {\bibfnamefont {I.}~\bibnamefont {Ga\ifmmode \check{s}\else
  \v{s}\fi{}pari\ifmmode~\acute{c}\else \'{c}\fi{}}}, \bibinfo {author}
  {\bibfnamefont {R.-B.}\ \bibnamefont {Gerst}}, \bibinfo {author}
  {\bibfnamefont {J.}~\bibnamefont {Gibelin}}, \bibinfo {author} {\bibfnamefont
  {K.~I.}\ \bibnamefont {Hahn}}, \bibinfo {author} {\bibfnamefont
  {D.}~\bibnamefont {Kim}}, \bibinfo {author} {\bibfnamefont {T.}~\bibnamefont
  {Koiwai}}, \bibinfo {author} {\bibfnamefont {Y.}~\bibnamefont {Kondo}},
  \bibinfo {author} {\bibfnamefont {P.}~\bibnamefont {Koseoglou}}, \bibinfo
  {author} {\bibfnamefont {J.}~\bibnamefont {Lee}}, \bibinfo {author}
  {\bibfnamefont {C.}~\bibnamefont {Lehr}}, \bibinfo {author} {\bibfnamefont
  {B.~D.}\ \bibnamefont {Linh}}, \bibinfo {author} {\bibfnamefont
  {T.}~\bibnamefont {Lokotko}}, \bibinfo {author} {\bibfnamefont
  {M.}~\bibnamefont {MacCormick}}, \bibinfo {author} {\bibfnamefont
  {K.}~\bibnamefont {Moschner}}, \bibinfo {author} {\bibfnamefont
  {T.}~\bibnamefont {Nakamura}}, \bibinfo {author} {\bibfnamefont {S.~Y.}\
  \bibnamefont {Park}}, \bibinfo {author} {\bibfnamefont {D.}~\bibnamefont
  {Rossi}}, \bibinfo {author} {\bibfnamefont {E.}~\bibnamefont {Sahin}},
  \bibinfo {author} {\bibfnamefont {D.}~\bibnamefont {Sohler}}, \bibinfo
  {author} {\bibfnamefont {P.-A.}\ \bibnamefont {S\"oderstr\"om}}, \bibinfo
  {author} {\bibfnamefont {S.}~\bibnamefont {Takeuchi}}, \bibinfo {author}
  {\bibfnamefont {H.}~\bibnamefont {T\"ornqvist}}, \bibinfo {author}
  {\bibfnamefont {V.}~\bibnamefont {Vaquero}}, \bibinfo {author} {\bibfnamefont
  {V.}~\bibnamefont {Wagner}}, \bibinfo {author} {\bibfnamefont
  {S.}~\bibnamefont {Wang}}, \bibinfo {author} {\bibfnamefont {V.}~\bibnamefont
  {Werner}}, \bibinfo {author} {\bibfnamefont {X.}~\bibnamefont {Xu}}, \bibinfo
  {author} {\bibfnamefont {H.}~\bibnamefont {Yamada}}, \bibinfo {author}
  {\bibfnamefont {D.}~\bibnamefont {Yan}}, \bibinfo {author} {\bibfnamefont
  {Z.}~\bibnamefont {Yang}}, \bibinfo {author} {\bibfnamefont {M.}~\bibnamefont
  {Yasuda}}, \ and\ \bibinfo {author} {\bibfnamefont {L.}~\bibnamefont
  {Zanetti}},\ }\bibfield  {title} {\enquote {\bibinfo {title} {How robust is
  the $n=34$ subshell closure? first spectroscopy of $^{52}\mathrm{Ar}$},}\
  }\href {\doibase 10.1103/PhysRevLett.122.072502} {\bibfield  {journal}
  {\bibinfo  {journal} {Phys. Rev. Lett.}\ }\textbf {\bibinfo {volume} {122}},\
  \bibinfo {pages} {072502} (\bibinfo {year} {2019})}\BibitemShut {NoStop}%
\bibitem [{\citenamefont {Taniuchi}\ \emph {et~al.}(2019)\citenamefont
  {Taniuchi}, \citenamefont {Santamaria}, \citenamefont {Doornenbal},
  \citenamefont {Obertelli}, \citenamefont {Yoneda}, \citenamefont {Authelet},
  \citenamefont {Baba}, \citenamefont {Calvet}, \citenamefont {Ch{\^{a}}teau},
  \citenamefont {Corsi}, \citenamefont {Delbart}, \citenamefont {Gheller},
  \citenamefont {Gillibert}, \citenamefont {Holt}, \citenamefont {Isobe},
  \citenamefont {Lapoux}, \citenamefont {Matsushita}, \citenamefont
  {Men{\'{e}}ndez}, \citenamefont {Momiyama}, \citenamefont {Motobayashi},
  \citenamefont {Niikura}, \citenamefont {Nowacki}, \citenamefont {Ogata},
  \citenamefont {Otsu}, \citenamefont {Otsuka}, \citenamefont {P{\'{e}}ron},
  \citenamefont {P{\'{e}}ru}, \citenamefont {Peyaud}, \citenamefont {Pollacco},
  \citenamefont {Poves}, \citenamefont {Rouss{\'{e}}}, \citenamefont {Sakurai},
  \citenamefont {Schwenk}, \citenamefont {Shiga}, \citenamefont {Simonis},
  \citenamefont {Stroberg}, \citenamefont {Takeuchi}, \citenamefont {Tsunoda},
  \citenamefont {Uesaka}, \citenamefont {Wang}, \citenamefont {Browne},
  \citenamefont {Chung}, \citenamefont {Dombradi}, \citenamefont {Franchoo},
  \citenamefont {Giacoppo}, \citenamefont {Gottardo}, \citenamefont
  {Hady{\'{n}}ska-Kl{\c{e}}k}, \citenamefont {Korkulu}, \citenamefont {Koyama},
  \citenamefont {Kubota}, \citenamefont {Lee}, \citenamefont {Lettmann},
  \citenamefont {Louchart}, \citenamefont {Lozeva}, \citenamefont {Matsui},
  \citenamefont {Miyazaki}, \citenamefont {Nishimura}, \citenamefont {Olivier},
  \citenamefont {Ota}, \citenamefont {Patel}, \citenamefont {Şahin},
  \citenamefont {Shand}, \citenamefont {S{\"{o}}derstr{\"{o}}m}, \citenamefont
  {Stefan}, \citenamefont {Steppenbeck}, \citenamefont {Sumikama},
  \citenamefont {Suzuki}, \citenamefont {Vajta}, \citenamefont {Werner},
  \citenamefont {Wu},\ and\ \citenamefont {Xu}}]{taniuchi2019}%
  \BibitemOpen
  \bibfield  {author} {\bibinfo {author} {\bibfnamefont {R}~\bibnamefont
  {Taniuchi}}, \bibinfo {author} {\bibfnamefont {C}~\bibnamefont {Santamaria}},
  \bibinfo {author} {\bibfnamefont {P}~\bibnamefont {Doornenbal}}, \bibinfo
  {author} {\bibfnamefont {A}~\bibnamefont {Obertelli}}, \bibinfo {author}
  {\bibfnamefont {K}~\bibnamefont {Yoneda}}, \bibinfo {author} {\bibfnamefont
  {G}~\bibnamefont {Authelet}}, \bibinfo {author} {\bibfnamefont
  {H}~\bibnamefont {Baba}}, \bibinfo {author} {\bibfnamefont {D}~\bibnamefont
  {Calvet}}, \bibinfo {author} {\bibfnamefont {F}~\bibnamefont
  {Ch{\^{a}}teau}}, \bibinfo {author} {\bibfnamefont {A}~\bibnamefont {Corsi}},
  \bibinfo {author} {\bibfnamefont {A}~\bibnamefont {Delbart}}, \bibinfo
  {author} {\bibfnamefont {J.-M.}\ \bibnamefont {Gheller}}, \bibinfo {author}
  {\bibfnamefont {A}~\bibnamefont {Gillibert}}, \bibinfo {author}
  {\bibfnamefont {J~D}\ \bibnamefont {Holt}}, \bibinfo {author} {\bibfnamefont
  {T}~\bibnamefont {Isobe}}, \bibinfo {author} {\bibfnamefont {V}~\bibnamefont
  {Lapoux}}, \bibinfo {author} {\bibfnamefont {M}~\bibnamefont {Matsushita}},
  \bibinfo {author} {\bibfnamefont {J}~\bibnamefont {Men{\'{e}}ndez}}, \bibinfo
  {author} {\bibfnamefont {S}~\bibnamefont {Momiyama}}, \bibinfo {author}
  {\bibfnamefont {T}~\bibnamefont {Motobayashi}}, \bibinfo {author}
  {\bibfnamefont {M}~\bibnamefont {Niikura}}, \bibinfo {author} {\bibfnamefont
  {F}~\bibnamefont {Nowacki}}, \bibinfo {author} {\bibfnamefont
  {K}~\bibnamefont {Ogata}}, \bibinfo {author} {\bibfnamefont {H}~\bibnamefont
  {Otsu}}, \bibinfo {author} {\bibfnamefont {T}~\bibnamefont {Otsuka}},
  \bibinfo {author} {\bibfnamefont {C}~\bibnamefont {P{\'{e}}ron}}, \bibinfo
  {author} {\bibfnamefont {S}~\bibnamefont {P{\'{e}}ru}}, \bibinfo {author}
  {\bibfnamefont {A}~\bibnamefont {Peyaud}}, \bibinfo {author} {\bibfnamefont
  {E~C}\ \bibnamefont {Pollacco}}, \bibinfo {author} {\bibfnamefont
  {A}~\bibnamefont {Poves}}, \bibinfo {author} {\bibfnamefont {J.-Y.}\
  \bibnamefont {Rouss{\'{e}}}}, \bibinfo {author} {\bibfnamefont
  {H}~\bibnamefont {Sakurai}}, \bibinfo {author} {\bibfnamefont
  {A}~\bibnamefont {Schwenk}}, \bibinfo {author} {\bibfnamefont
  {Y}~\bibnamefont {Shiga}}, \bibinfo {author} {\bibfnamefont {J}~\bibnamefont
  {Simonis}}, \bibinfo {author} {\bibfnamefont {S~R}\ \bibnamefont {Stroberg}},
  \bibinfo {author} {\bibfnamefont {S}~\bibnamefont {Takeuchi}}, \bibinfo
  {author} {\bibfnamefont {Y}~\bibnamefont {Tsunoda}}, \bibinfo {author}
  {\bibfnamefont {T}~\bibnamefont {Uesaka}}, \bibinfo {author} {\bibfnamefont
  {H}~\bibnamefont {Wang}}, \bibinfo {author} {\bibfnamefont {F}~\bibnamefont
  {Browne}}, \bibinfo {author} {\bibfnamefont {L~X}\ \bibnamefont {Chung}},
  \bibinfo {author} {\bibfnamefont {Z}~\bibnamefont {Dombradi}}, \bibinfo
  {author} {\bibfnamefont {S}~\bibnamefont {Franchoo}}, \bibinfo {author}
  {\bibfnamefont {F}~\bibnamefont {Giacoppo}}, \bibinfo {author} {\bibfnamefont
  {A}~\bibnamefont {Gottardo}}, \bibinfo {author} {\bibfnamefont
  {K}~\bibnamefont {Hady{\'{n}}ska-Kl{\c{e}}k}}, \bibinfo {author}
  {\bibfnamefont {Z}~\bibnamefont {Korkulu}}, \bibinfo {author} {\bibfnamefont
  {S}~\bibnamefont {Koyama}}, \bibinfo {author} {\bibfnamefont {Y}~\bibnamefont
  {Kubota}}, \bibinfo {author} {\bibfnamefont {J}~\bibnamefont {Lee}}, \bibinfo
  {author} {\bibfnamefont {M}~\bibnamefont {Lettmann}}, \bibinfo {author}
  {\bibfnamefont {C}~\bibnamefont {Louchart}}, \bibinfo {author} {\bibfnamefont
  {R}~\bibnamefont {Lozeva}}, \bibinfo {author} {\bibfnamefont {K}~\bibnamefont
  {Matsui}}, \bibinfo {author} {\bibfnamefont {T}~\bibnamefont {Miyazaki}},
  \bibinfo {author} {\bibfnamefont {S}~\bibnamefont {Nishimura}}, \bibinfo
  {author} {\bibfnamefont {L}~\bibnamefont {Olivier}}, \bibinfo {author}
  {\bibfnamefont {S}~\bibnamefont {Ota}}, \bibinfo {author} {\bibfnamefont
  {Z}~\bibnamefont {Patel}}, \bibinfo {author} {\bibfnamefont {E}~\bibnamefont
  {Şahin}}, \bibinfo {author} {\bibfnamefont {C}~\bibnamefont {Shand}},
  \bibinfo {author} {\bibfnamefont {P.-A.}\ \bibnamefont
  {S{\"{o}}derstr{\"{o}}m}}, \bibinfo {author} {\bibfnamefont {I}~\bibnamefont
  {Stefan}}, \bibinfo {author} {\bibfnamefont {D}~\bibnamefont {Steppenbeck}},
  \bibinfo {author} {\bibfnamefont {T}~\bibnamefont {Sumikama}}, \bibinfo
  {author} {\bibfnamefont {D}~\bibnamefont {Suzuki}}, \bibinfo {author}
  {\bibfnamefont {Z}~\bibnamefont {Vajta}}, \bibinfo {author} {\bibfnamefont
  {V}~\bibnamefont {Werner}}, \bibinfo {author} {\bibfnamefont {J}~\bibnamefont
  {Wu}}, \ and\ \bibinfo {author} {\bibfnamefont {Z~Y}\ \bibnamefont {Xu}},\
  }\bibfield  {title} {\enquote {\bibinfo {title} {{78Ni revealed as a doubly
  magic stronghold against nuclear deformation}},}\ }\href {\doibase
  10.1038/s41586-019-1155-x} {\bibfield  {journal} {\bibinfo  {journal}
  {Nature}\ }\textbf {\bibinfo {volume} {569}},\ \bibinfo {pages} {53--58}
  (\bibinfo {year} {2019})}\BibitemShut {NoStop}%
\bibitem [{\citenamefont {Gysbers}\ \emph {et~al.}(2019)\citenamefont
  {Gysbers}, \citenamefont {Hagen}, \citenamefont {Holt}, \citenamefont
  {Jansen}, \citenamefont {Morris}, \citenamefont {Navr{\'a}til}, \citenamefont
  {Papenbrock}, \citenamefont {Quaglioni}, \citenamefont {Schwenk},
  \citenamefont {Stroberg},\ and\ \citenamefont {Wendt}}]{Gysbers2019}%
  \BibitemOpen
  \bibfield  {author} {\bibinfo {author} {\bibfnamefont {P.}~\bibnamefont
  {Gysbers}}, \bibinfo {author} {\bibfnamefont {G.}~\bibnamefont {Hagen}},
  \bibinfo {author} {\bibfnamefont {J.~D.}\ \bibnamefont {Holt}}, \bibinfo
  {author} {\bibfnamefont {G.~R.}\ \bibnamefont {Jansen}}, \bibinfo {author}
  {\bibfnamefont {T.~D.}\ \bibnamefont {Morris}}, \bibinfo {author}
  {\bibfnamefont {P.}~\bibnamefont {Navr{\'a}til}}, \bibinfo {author}
  {\bibfnamefont {T.}~\bibnamefont {Papenbrock}}, \bibinfo {author}
  {\bibfnamefont {S.}~\bibnamefont {Quaglioni}}, \bibinfo {author}
  {\bibfnamefont {A.}~\bibnamefont {Schwenk}}, \bibinfo {author} {\bibfnamefont
  {S.~R.}\ \bibnamefont {Stroberg}}, \ and\ \bibinfo {author} {\bibfnamefont
  {K.~A.}\ \bibnamefont {Wendt}},\ }\bibfield  {title} {\enquote {\bibinfo
  {title} {Discrepancy between experimental and theoretical b-decay rates
  resolved from first principles},}\ }\href {\doibase
  10.1038/s41567-019-0450-7} {\bibfield  {journal} {\bibinfo  {journal} {Nature
  Physics}\ } (\bibinfo {year} {2019}),\ 10.1038/s41567-019-0450-7}\BibitemShut
  {NoStop}%
\bibitem [{\citenamefont {{Holt}}\ \emph {et~al.}(2019)\citenamefont {{Holt}},
  \citenamefont {{Stroberg}}, \citenamefont {{Schwenk}},\ and\ \citenamefont
  {{Simonis}}}]{holt2019}%
  \BibitemOpen
  \bibfield  {author} {\bibinfo {author} {\bibfnamefont {J.~D.}\ \bibnamefont
  {{Holt}}}, \bibinfo {author} {\bibfnamefont {S.~R.}\ \bibnamefont
  {{Stroberg}}}, \bibinfo {author} {\bibfnamefont {A.}~\bibnamefont
  {{Schwenk}}}, \ and\ \bibinfo {author} {\bibfnamefont {J.}~\bibnamefont
  {{Simonis}}},\ }\bibfield  {title} {\enquote {\bibinfo {title} {{Ab initio
  limits of atomic nuclei}},}\ }\href@noop {} {\bibfield  {journal} {\bibinfo
  {journal} {arXiv e-prints}\ ,\ \bibinfo {eid} {arXiv:1905.10475}} (\bibinfo
  {year} {2019})},\ \Eprint {http://arxiv.org/abs/1905.10475} {arXiv:1905.10475
  [nucl-th]} \BibitemShut {NoStop}%
\bibitem [{\citenamefont {{Som{\`a}}}\ \emph {et~al.}(2019)\citenamefont
  {{Som{\`a}}}, \citenamefont {{Navr{\'a}til}}, \citenamefont {{Raimondi}},
  \citenamefont {{Barbieri}},\ and\ \citenamefont {{Duguet}}}]{soma2019}%
  \BibitemOpen
  \bibfield  {author} {\bibinfo {author} {\bibfnamefont {V.}~\bibnamefont
  {{Som{\`a}}}}, \bibinfo {author} {\bibfnamefont {P.}~\bibnamefont
  {{Navr{\'a}til}}}, \bibinfo {author} {\bibfnamefont {F.}~\bibnamefont
  {{Raimondi}}}, \bibinfo {author} {\bibfnamefont {C.}~\bibnamefont
  {{Barbieri}}}, \ and\ \bibinfo {author} {\bibfnamefont {T.}~\bibnamefont
  {{Duguet}}},\ }\bibfield  {title} {\enquote {\bibinfo {title} {{Novel chiral
  Hamiltonian and observables in light and medium-mass nuclei}},}\ }\href@noop
  {} {\bibfield  {journal} {\bibinfo  {journal} {arXiv e-prints}\ ,\ \bibinfo
  {eid} {arXiv:1907.09790}} (\bibinfo {year} {2019})},\ \Eprint
  {http://arxiv.org/abs/1907.09790} {arXiv:1907.09790 [nucl-th]} \BibitemShut
  {NoStop}%
\bibitem [{\citenamefont {Kolck}(1999)}]{vankolck1999}%
  \BibitemOpen
  \bibfield  {author} {\bibinfo {author} {\bibfnamefont {U.~Van}\ \bibnamefont
  {Kolck}},\ }\bibfield  {title} {\enquote {\bibinfo {title} {Effective field
  theory of nuclear forces},}\ }\href {\doibase 10.1016/S0146-6410(99)00097-6}
  {\bibfield  {journal} {\bibinfo  {journal} {Prog. Part. Nucl. Phys.}\
  }\textbf {\bibinfo {volume} {43}},\ \bibinfo {pages} {337 -- 418} (\bibinfo
  {year} {1999})}\BibitemShut {NoStop}%
\bibitem [{\citenamefont {Epelbaum}\ \emph {et~al.}(2009)\citenamefont
  {Epelbaum}, \citenamefont {Hammer},\ and\ \citenamefont
  {Mei\ss{}ner}}]{epelbaum2009}%
  \BibitemOpen
  \bibfield  {author} {\bibinfo {author} {\bibfnamefont {E.}~\bibnamefont
  {Epelbaum}}, \bibinfo {author} {\bibfnamefont {H.-W.}\ \bibnamefont
  {Hammer}}, \ and\ \bibinfo {author} {\bibfnamefont {U.-G.}\ \bibnamefont
  {Mei\ss{}ner}},\ }\bibfield  {title} {\enquote {\bibinfo {title} {Modern
  theory of nuclear forces},}\ }\href {\doibase 10.1103/RevModPhys.81.1773}
  {\bibfield  {journal} {\bibinfo  {journal} {Rev. Mod. Phys.}\ }\textbf
  {\bibinfo {volume} {81}},\ \bibinfo {pages} {1773--1825} (\bibinfo {year}
  {2009})}\BibitemShut {NoStop}%
\bibitem [{\citenamefont {Machleidt}\ and\ \citenamefont
  {Entem}(2011)}]{machleidt2011}%
  \BibitemOpen
  \bibfield  {author} {\bibinfo {author} {\bibfnamefont {R.}~\bibnamefont
  {Machleidt}}\ and\ \bibinfo {author} {\bibfnamefont {D.~R.}\ \bibnamefont
  {Entem}},\ }\bibfield  {title} {\enquote {\bibinfo {title} {Chiral effective
  field theory and nuclear forces},}\ }\href {\doibase
  10.1016/j.physrep.2011.02.001} {\bibfield  {journal} {\bibinfo  {journal}
  {Phys. Rep.}\ }\textbf {\bibinfo {volume} {503}},\ \bibinfo {pages} {1 -- 75}
  (\bibinfo {year} {2011})}\BibitemShut {NoStop}%
\bibitem [{\citenamefont {Ekstr\"om}\ \emph {et~al.}(2015)\citenamefont
  {Ekstr\"om}, \citenamefont {Jansen}, \citenamefont {Wendt}, \citenamefont
  {Hagen}, \citenamefont {Papenbrock}, \citenamefont {Carlsson}, \citenamefont
  {Forss\'en}, \citenamefont {Hjorth-Jensen}, \citenamefont {Navr\'atil},\ and\
  \citenamefont {Nazarewicz}}]{ekstrom2015}%
  \BibitemOpen
  \bibfield  {author} {\bibinfo {author} {\bibfnamefont {A.}~\bibnamefont
  {Ekstr\"om}}, \bibinfo {author} {\bibfnamefont {G.~R.}\ \bibnamefont
  {Jansen}}, \bibinfo {author} {\bibfnamefont {K.~A.}\ \bibnamefont {Wendt}},
  \bibinfo {author} {\bibfnamefont {G.}~\bibnamefont {Hagen}}, \bibinfo
  {author} {\bibfnamefont {T.}~\bibnamefont {Papenbrock}}, \bibinfo {author}
  {\bibfnamefont {B.~D.}\ \bibnamefont {Carlsson}}, \bibinfo {author}
  {\bibfnamefont {C.}~\bibnamefont {Forss\'en}}, \bibinfo {author}
  {\bibfnamefont {M.}~\bibnamefont {Hjorth-Jensen}}, \bibinfo {author}
  {\bibfnamefont {P.}~\bibnamefont {Navr\'atil}}, \ and\ \bibinfo {author}
  {\bibfnamefont {W.}~\bibnamefont {Nazarewicz}},\ }\bibfield  {title}
  {\enquote {\bibinfo {title} {Accurate nuclear radii and binding energies from
  a chiral interaction},}\ }\href {\doibase 10.1103/PhysRevC.91.051301}
  {\bibfield  {journal} {\bibinfo  {journal} {Phys. Rev. C}\ }\textbf {\bibinfo
  {volume} {91}},\ \bibinfo {pages} {051301} (\bibinfo {year}
  {2015})}\BibitemShut {NoStop}%
\bibitem [{\citenamefont {Nogga}\ \emph {et~al.}(2004)\citenamefont {Nogga},
  \citenamefont {Bogner},\ and\ \citenamefont {Schwenk}}]{nogga2004}%
  \BibitemOpen
  \bibfield  {author} {\bibinfo {author} {\bibfnamefont {A.}~\bibnamefont
  {Nogga}}, \bibinfo {author} {\bibfnamefont {S.~K.}\ \bibnamefont {Bogner}}, \
  and\ \bibinfo {author} {\bibfnamefont {A.}~\bibnamefont {Schwenk}},\
  }\bibfield  {title} {\enquote {\bibinfo {title} {Low-momentum interaction in
  few-nucleon systems},}\ }\href {\doibase 10.1103/PhysRevC.70.061002}
  {\bibfield  {journal} {\bibinfo  {journal} {Phys. Rev. C}\ }\textbf {\bibinfo
  {volume} {70}},\ \bibinfo {pages} {061002} (\bibinfo {year}
  {2004})}\BibitemShut {NoStop}%
\bibitem [{\citenamefont {Hebeler}\ \emph {et~al.}(2011)\citenamefont
  {Hebeler}, \citenamefont {Bogner}, \citenamefont {Furnstahl}, \citenamefont
  {Nogga},\ and\ \citenamefont {Schwenk}}]{hebeler2011}%
  \BibitemOpen
  \bibfield  {author} {\bibinfo {author} {\bibfnamefont {K.}~\bibnamefont
  {Hebeler}}, \bibinfo {author} {\bibfnamefont {S.~K.}\ \bibnamefont {Bogner}},
  \bibinfo {author} {\bibfnamefont {R.~J.}\ \bibnamefont {Furnstahl}}, \bibinfo
  {author} {\bibfnamefont {A.}~\bibnamefont {Nogga}}, \ and\ \bibinfo {author}
  {\bibfnamefont {A.}~\bibnamefont {Schwenk}},\ }\bibfield  {title} {\enquote
  {\bibinfo {title} {Improved nuclear matter calculations from chiral
  low-momentum interactions},}\ }\href {\doibase 10.1103/PhysRevC.83.031301}
  {\bibfield  {journal} {\bibinfo  {journal} {Phys. Rev. C}\ }\textbf {\bibinfo
  {volume} {83}},\ \bibinfo {pages} {031301} (\bibinfo {year}
  {2011})}\BibitemShut {NoStop}%
\bibitem [{\citenamefont {Schindler}\ and\ \citenamefont
  {Phillips}(2009)}]{Schindler2009}%
  \BibitemOpen
  \bibfield  {author} {\bibinfo {author} {\bibfnamefont {M.R.}\ \bibnamefont
  {Schindler}}\ and\ \bibinfo {author} {\bibfnamefont {D.R.}\ \bibnamefont
  {Phillips}},\ }\bibfield  {title} {\enquote {\bibinfo {title} {Bayesian
  methods for parameter estimation in effective field theories},}\ }\href
  {\doibase 10.1016/j.aop.2008.09.003} {\bibfield  {journal} {\bibinfo
  {journal} {Ann. Phys.}\ }\textbf {\bibinfo {volume} {324}},\ \bibinfo {pages}
  {682 -- 708} (\bibinfo {year} {2009})}\BibitemShut {NoStop}%
\bibitem [{\citenamefont {Wesolowski}\ \emph {et~al.}(2016)\citenamefont
  {Wesolowski}, \citenamefont {Klco}, \citenamefont {Furnstahl}, \citenamefont
  {Phillips},\ and\ \citenamefont {liya}}]{Wesolowski2015}%
  \BibitemOpen
  \bibfield  {author} {\bibinfo {author} {\bibfnamefont {S}~\bibnamefont
  {Wesolowski}}, \bibinfo {author} {\bibfnamefont {N}~\bibnamefont {Klco}},
  \bibinfo {author} {\bibfnamefont {R~J}\ \bibnamefont {Furnstahl}}, \bibinfo
  {author} {\bibfnamefont {D~R}\ \bibnamefont {Phillips}}, \ and\ \bibinfo
  {author} {\bibfnamefont {A~Thapa}\ \bibnamefont {liya}},\ }\bibfield  {title}
  {\enquote {\bibinfo {title} {Bayesian parameter estimation for effective
  field theories},}\ }\href {http://stacks.iop.org/0954-3899/43/i=7/a=074001}
  {\bibfield  {journal} {\bibinfo  {journal} {Journal of Physics G: Nuclear and
  Particle Physics}\ }\textbf {\bibinfo {volume} {43}},\ \bibinfo {pages}
  {074001} (\bibinfo {year} {2016})}\BibitemShut {NoStop}%
\bibitem [{\citenamefont {Wesolowski}\ \emph {et~al.}(2019)\citenamefont
  {Wesolowski}, \citenamefont {Furnstahl}, \citenamefont {Melendez},\ and\
  \citenamefont {Phillips}}]{Wesolowski2019}%
  \BibitemOpen
  \bibfield  {author} {\bibinfo {author} {\bibfnamefont {S}~\bibnamefont
  {Wesolowski}}, \bibinfo {author} {\bibfnamefont {R~J}\ \bibnamefont
  {Furnstahl}}, \bibinfo {author} {\bibfnamefont {J~A}\ \bibnamefont
  {Melendez}}, \ and\ \bibinfo {author} {\bibfnamefont {D~R}\ \bibnamefont
  {Phillips}},\ }\bibfield  {title} {\enquote {\bibinfo {title} {Exploring
  bayesian parameter estimation for chiral effective field theory using
  nucleon{\textendash}nucleon phase shifts},}\ }\href {\doibase
  10.1088/1361-6471/aaf5fc} {\bibfield  {journal} {\bibinfo  {journal} {Journal
  of Physics G: Nuclear and Particle Physics}\ }\textbf {\bibinfo {volume}
  {46}},\ \bibinfo {pages} {045102} (\bibinfo {year} {2019})}\BibitemShut
  {NoStop}%
\bibitem [{\citenamefont {Frame}\ \emph {et~al.}(2018)\citenamefont {Frame},
  \citenamefont {He}, \citenamefont {Ipsen}, \citenamefont {Lee}, \citenamefont
  {Lee},\ and\ \citenamefont {Rrapaj}}]{Frame2018}%
  \BibitemOpen
  \bibfield  {author} {\bibinfo {author} {\bibfnamefont {Dillon}\ \bibnamefont
  {Frame}}, \bibinfo {author} {\bibfnamefont {Rongzheng}\ \bibnamefont {He}},
  \bibinfo {author} {\bibfnamefont {Ilse}\ \bibnamefont {Ipsen}}, \bibinfo
  {author} {\bibfnamefont {Daniel}\ \bibnamefont {Lee}}, \bibinfo {author}
  {\bibfnamefont {Dean}\ \bibnamefont {Lee}}, \ and\ \bibinfo {author}
  {\bibfnamefont {Ermal}\ \bibnamefont {Rrapaj}},\ }\bibfield  {title}
  {\enquote {\bibinfo {title} {Eigenvector continuation with subspace
  learning},}\ }\href {\doibase 10.1103/PhysRevLett.121.032501} {\bibfield
  {journal} {\bibinfo  {journal} {Phys. Rev. Lett.}\ }\textbf {\bibinfo
  {volume} {121}},\ \bibinfo {pages} {032501} (\bibinfo {year}
  {2018})}\BibitemShut {NoStop}%
\bibitem [{\citenamefont {Coester}(1958)}]{coester1958}%
  \BibitemOpen
  \bibfield  {author} {\bibinfo {author} {\bibfnamefont {F.}~\bibnamefont
  {Coester}},\ }\bibfield  {title} {\enquote {\bibinfo {title} {Bound states of
  a many-particle system},}\ }\href {\doibase 10.1016/0029-5582(58)90280-3}
  {\bibfield  {journal} {\bibinfo  {journal} {Nucl. Phys.}\ }\textbf {\bibinfo
  {volume} {7}},\ \bibinfo {pages} {421 -- 424} (\bibinfo {year}
  {1958})}\BibitemShut {NoStop}%
\bibitem [{\citenamefont {Coester}\ and\ \citenamefont
  {K{\"u}mmel}(1960)}]{coester1960}%
  \BibitemOpen
  \bibfield  {author} {\bibinfo {author} {\bibfnamefont {F.}~\bibnamefont
  {Coester}}\ and\ \bibinfo {author} {\bibfnamefont {H.}~\bibnamefont
  {K{\"u}mmel}},\ }\bibfield  {title} {\enquote {\bibinfo {title} {Short-range
  correlations in nuclear wave functions},}\ }\href {\doibase
  10.1016/0029-5582(60)90140-1} {\bibfield  {journal} {\bibinfo  {journal}
  {Nucl. Phys.}\ }\textbf {\bibinfo {volume} {17}},\ \bibinfo {pages} {477 --
  485} (\bibinfo {year} {1960})}\BibitemShut {NoStop}%
\bibitem [{\citenamefont {K{\"u}mmel}\ \emph {et~al.}(1978)\citenamefont
  {K{\"u}mmel}, \citenamefont {L{\"u}hrmann},\ and\ \citenamefont
  {Zabolitzky}}]{kuemmel1978}%
  \BibitemOpen
  \bibfield  {author} {\bibinfo {author} {\bibfnamefont {H.}~\bibnamefont
  {K{\"u}mmel}}, \bibinfo {author} {\bibfnamefont {K.~H.}\ \bibnamefont
  {L{\"u}hrmann}}, \ and\ \bibinfo {author} {\bibfnamefont {J.~G.}\
  \bibnamefont {Zabolitzky}},\ }\bibfield  {title} {\enquote {\bibinfo {title}
  {{Many-fermion theory in exp S- (or coupled cluster) form}},}\ }\href
  {\doibase 10.1016/0370-1573(78)90081-9} {\bibfield  {journal} {\bibinfo
  {journal} {Phys. Rep.}\ }\textbf {\bibinfo {volume} {36}},\ \bibinfo {pages}
  {1 -- 63} (\bibinfo {year} {1978})}\BibitemShut {NoStop}%
\bibitem [{\citenamefont {Mihaila}\ and\ \citenamefont
  {Heisenberg}(2000)}]{mihaila2000b}%
  \BibitemOpen
  \bibfield  {author} {\bibinfo {author} {\bibfnamefont {B.}~\bibnamefont
  {Mihaila}}\ and\ \bibinfo {author} {\bibfnamefont {J.~H.}\ \bibnamefont
  {Heisenberg}},\ }\bibfield  {title} {\enquote {\bibinfo {title} {{Microscopic
  Calculation of the Inclusive Electron Scattering Structure Function in
  $^{16}$O}},}\ }\href {\doibase 10.1103/PhysRevLett.84.1403} {\bibfield
  {journal} {\bibinfo  {journal} {Phys. Rev. Lett.}\ }\textbf {\bibinfo
  {volume} {84}},\ \bibinfo {pages} {1403--1406} (\bibinfo {year}
  {2000})}\BibitemShut {NoStop}%
\bibitem [{\citenamefont {Dean}\ and\ \citenamefont
  {Hjorth-Jensen}(2004)}]{dean2004}%
  \BibitemOpen
  \bibfield  {author} {\bibinfo {author} {\bibfnamefont {D.~J.}\ \bibnamefont
  {Dean}}\ and\ \bibinfo {author} {\bibfnamefont {M.}~\bibnamefont
  {Hjorth-Jensen}},\ }\bibfield  {title} {\enquote {\bibinfo {title}
  {Coupled-cluster approach to nuclear physics},}\ }\href {\doibase
  10.1103/PhysRevC.69.054320} {\bibfield  {journal} {\bibinfo  {journal} {Phys.
  Rev. C}\ }\textbf {\bibinfo {volume} {69}},\ \bibinfo {pages} {054320}
  (\bibinfo {year} {2004})}\BibitemShut {NoStop}%
\bibitem [{\citenamefont {Bartlett}\ and\ \citenamefont
  {Musia\l{}}(2007)}]{bartlett2007}%
  \BibitemOpen
  \bibfield  {author} {\bibinfo {author} {\bibfnamefont {R.~J.}\ \bibnamefont
  {Bartlett}}\ and\ \bibinfo {author} {\bibfnamefont {M.}~\bibnamefont
  {Musia\l{}}},\ }\bibfield  {title} {\enquote {\bibinfo {title}
  {Coupled-cluster theory in quantum chemistry},}\ }\href {\doibase
  10.1103/RevModPhys.79.291} {\bibfield  {journal} {\bibinfo  {journal} {Rev.
  Mod. Phys.}\ }\textbf {\bibinfo {volume} {79}},\ \bibinfo {pages} {291--352}
  (\bibinfo {year} {2007})}\BibitemShut {NoStop}%
\bibitem [{\citenamefont {Hagen}\ \emph {et~al.}(2014)\citenamefont {Hagen},
  \citenamefont {Papenbrock}, \citenamefont {Hjorth-Jensen},\ and\
  \citenamefont {Dean}}]{hagen2014}%
  \BibitemOpen
  \bibfield  {author} {\bibinfo {author} {\bibfnamefont {G.}~\bibnamefont
  {Hagen}}, \bibinfo {author} {\bibfnamefont {T.}~\bibnamefont {Papenbrock}},
  \bibinfo {author} {\bibfnamefont {M.}~\bibnamefont {Hjorth-Jensen}}, \ and\
  \bibinfo {author} {\bibfnamefont {D.~J.}\ \bibnamefont {Dean}},\ }\bibfield
  {title} {\enquote {\bibinfo {title} {Coupled-cluster computations of atomic
  nuclei},}\ }\href {\doibase 10.1088/0034-4885/77/9/096302} {\bibfield
  {journal} {\bibinfo  {journal} {Rep. Prog. Phys.}\ }\textbf {\bibinfo
  {volume} {77}},\ \bibinfo {pages} {096302} (\bibinfo {year}
  {2014})}\BibitemShut {NoStop}%
\bibitem [{\citenamefont {{K{\"o}nig}}\ \emph {et~al.}(2019)\citenamefont
  {{K{\"o}nig}}, \citenamefont {{Ekstr{\"o}m}}, \citenamefont {{Hebeler}},
  \citenamefont {{Lee}},\ and\ \citenamefont {{Schwenk}}}]{konig2019}%
  \BibitemOpen
  \bibfield  {author} {\bibinfo {author} {\bibfnamefont {S.}~\bibnamefont
  {{K{\"o}nig}}}, \bibinfo {author} {\bibfnamefont {A.}~\bibnamefont
  {{Ekstr{\"o}m}}}, \bibinfo {author} {\bibfnamefont {K.}~\bibnamefont
  {{Hebeler}}}, \bibinfo {author} {\bibfnamefont {D.}~\bibnamefont {{Lee}}}, \
  and\ \bibinfo {author} {\bibfnamefont {A.}~\bibnamefont {{Schwenk}}},\
  }\bibfield  {title} {\enquote {\bibinfo {title} {{Eigenvector Continuation as
  an Efficient and Accurate Emulator for Uncertainty Quantification}},}\
  }\href@noop {} {\bibfield  {journal} {\bibinfo  {journal} {arXiv e-prints}\
  ,\ \bibinfo {eid} {arXiv:1909.08446}} (\bibinfo {year} {2019})},\ \Eprint
  {http://arxiv.org/abs/1909.08446} {arXiv:1909.08446 [nucl-th]} \BibitemShut
  {NoStop}%
\bibitem [{\citenamefont {{Ekstr{\"o}m}}\ \emph {et~al.}(2017)\citenamefont
  {{Ekstr{\"o}m}}, \citenamefont {{Hagen}}, \citenamefont {{Morris}},
  \citenamefont {{Papenbrock}},\ and\ \citenamefont
  {{Schwartz}}}]{ekstrom2017}%
  \BibitemOpen
  \bibfield  {author} {\bibinfo {author} {\bibfnamefont {A.}~\bibnamefont
  {{Ekstr{\"o}m}}}, \bibinfo {author} {\bibfnamefont {G.}~\bibnamefont
  {{Hagen}}}, \bibinfo {author} {\bibfnamefont {T.~D.}\ \bibnamefont
  {{Morris}}}, \bibinfo {author} {\bibfnamefont {T.}~\bibnamefont
  {{Papenbrock}}}, \ and\ \bibinfo {author} {\bibfnamefont {P.~D.}\
  \bibnamefont {{Schwartz}}},\ }\bibfield  {title} {\enquote {\bibinfo {title}
  {{Delta isobars and nuclear saturation}},}\ }\href@noop {} {\bibfield
  {journal} {\bibinfo  {journal} {ArXiv e-prints}\ } (\bibinfo {year}
  {2017})},\ \Eprint {http://arxiv.org/abs/1707.09028} {arXiv:1707.09028
  [nucl-th]} \BibitemShut {NoStop}%
\bibitem [{\citenamefont {Drischler}\ \emph {et~al.}(2019)\citenamefont
  {Drischler}, \citenamefont {Hebeler},\ and\ \citenamefont
  {Schwenk}}]{drischler2019}%
  \BibitemOpen
  \bibfield  {author} {\bibinfo {author} {\bibfnamefont {C.}~\bibnamefont
  {Drischler}}, \bibinfo {author} {\bibfnamefont {K.}~\bibnamefont {Hebeler}},
  \ and\ \bibinfo {author} {\bibfnamefont {A.}~\bibnamefont {Schwenk}},\
  }\bibfield  {title} {\enquote {\bibinfo {title} {Chiral interactions up to
  next-to-next-to-next-to-leading order and nuclear saturation},}\ }\href
  {\doibase 10.1103/PhysRevLett.122.042501} {\bibfield  {journal} {\bibinfo
  {journal} {Phys. Rev. Lett.}\ }\textbf {\bibinfo {volume} {122}},\ \bibinfo
  {pages} {042501} (\bibinfo {year} {2019})}\BibitemShut {NoStop}%
\bibitem [{\citenamefont {Jim\'enez-Hoyos}\ \emph {et~al.}(2012)\citenamefont
  {Jim\'enez-Hoyos}, \citenamefont {Rodr\'{\i}guez-Guzm\'an},\ and\
  \citenamefont {Scuseria}}]{hoyos2012}%
  \BibitemOpen
  \bibfield  {author} {\bibinfo {author} {\bibfnamefont {Carlos~A.}\
  \bibnamefont {Jim\'enez-Hoyos}}, \bibinfo {author} {\bibfnamefont
  {R.}~\bibnamefont {Rodr\'{\i}guez-Guzm\'an}}, \ and\ \bibinfo {author}
  {\bibfnamefont {Gustavo~E.}\ \bibnamefont {Scuseria}},\ }\bibfield  {title}
  {\enquote {\bibinfo {title} {$n$-electron slater determinants from nonunitary
  canonical transformations of fermion operators},}\ }\href {\doibase
  10.1103/PhysRevA.86.052102} {\bibfield  {journal} {\bibinfo  {journal} {Phys.
  Rev. A}\ }\textbf {\bibinfo {volume} {86}},\ \bibinfo {pages} {052102}
  (\bibinfo {year} {2012})}\BibitemShut {NoStop}%
\bibitem [{\citenamefont {Plasser}\ \emph {et~al.}(2016)\citenamefont
  {Plasser}, \citenamefont {Ruckenbauer}, \citenamefont {Mai}, \citenamefont
  {Oppel}, \citenamefont {Marquetand},\ and\ \citenamefont
  {González}}]{plasser2016}%
  \BibitemOpen
  \bibfield  {author} {\bibinfo {author} {\bibfnamefont {Felix}\ \bibnamefont
  {Plasser}}, \bibinfo {author} {\bibfnamefont {Matthias}\ \bibnamefont
  {Ruckenbauer}}, \bibinfo {author} {\bibfnamefont {Sebastian}\ \bibnamefont
  {Mai}}, \bibinfo {author} {\bibfnamefont {Markus}\ \bibnamefont {Oppel}},
  \bibinfo {author} {\bibfnamefont {Philipp}\ \bibnamefont {Marquetand}}, \
  and\ \bibinfo {author} {\bibfnamefont {Leticia}\ \bibnamefont {González}},\
  }\bibfield  {title} {\enquote {\bibinfo {title} {Efficient and flexible
  computation of many-electron wave function overlaps},}\ }\href {\doibase
  10.1021/acs.jctc.5b01148} {\bibfield  {journal} {\bibinfo  {journal} {Journal
  of Chemical Theory and Computation}\ }\textbf {\bibinfo {volume} {12}},\
  \bibinfo {pages} {1207--1219} (\bibinfo {year} {2016})},\ \bibinfo {note}
  {pMID: 26854874},\ \Eprint
  {http://arxiv.org/abs/https://doi.org/10.1021/acs.jctc.5b01148}
  {https://doi.org/10.1021/acs.jctc.5b01148} \BibitemShut {NoStop}%
\bibitem [{\citenamefont {Sobolâ²}(2001)}]{Sobol}%
  \BibitemOpen
  \bibfield  {author} {\bibinfo {author} {\bibfnamefont {I.M}\ \bibnamefont
  {Sobolâ²}},\ }\bibfield  {title} {\enquote {\bibinfo {title} {Global
  sensitivity indices for nonlinear mathematical models and their monte carlo
  estimates},}\ }\href {\doibase https://doi.org/10.1016/S0378-4754(00)00270-6}
  {\bibfield  {journal} {\bibinfo  {journal} {Mathematics and Computers in
  Simulation}\ }\textbf {\bibinfo {volume} {55}},\ \bibinfo {pages} {271 --
  280} (\bibinfo {year} {2001})},\ \bibinfo {note} {the Second IMACS Seminar on
  Monte Carlo Methods}\BibitemShut {NoStop}%
\bibitem [{\citenamefont {Saltelli}\ \emph {et~al.}(2010)\citenamefont
  {Saltelli}, \citenamefont {Annoni}, \citenamefont {Azzini}, \citenamefont
  {Campolongo}, \citenamefont {Ratto},\ and\ \citenamefont
  {Tarantola}}]{Saltelli}%
  \BibitemOpen
  \bibfield  {author} {\bibinfo {author} {\bibfnamefont {Andrea}\ \bibnamefont
  {Saltelli}}, \bibinfo {author} {\bibfnamefont {Paola}\ \bibnamefont
  {Annoni}}, \bibinfo {author} {\bibfnamefont {Ivano}\ \bibnamefont {Azzini}},
  \bibinfo {author} {\bibfnamefont {Francesca}\ \bibnamefont {Campolongo}},
  \bibinfo {author} {\bibfnamefont {Marco}\ \bibnamefont {Ratto}}, \ and\
  \bibinfo {author} {\bibfnamefont {Stefano}\ \bibnamefont {Tarantola}},\
  }\bibfield  {title} {\enquote {\bibinfo {title} {Variance based sensitivity
  analysis of model output. design and estimator for the total sensitivity
  index},}\ }\href {\doibase https://doi.org/10.1016/j.cpc.2009.09.018}
  {\bibfield  {journal} {\bibinfo  {journal} {Computer Physics Communications}\
  }\textbf {\bibinfo {volume} {181}},\ \bibinfo {pages} {259 -- 270} (\bibinfo
  {year} {2010})}\BibitemShut {NoStop}%
\bibitem [{\citenamefont {Homma}\ and\ \citenamefont {Saltelli}(1996)}]{Homma}%
  \BibitemOpen
  \bibfield  {author} {\bibinfo {author} {\bibfnamefont {Toshimitsu}\
  \bibnamefont {Homma}}\ and\ \bibinfo {author} {\bibfnamefont {Andrea}\
  \bibnamefont {Saltelli}},\ }\bibfield  {title} {\enquote {\bibinfo {title}
  {Importance measures in global sensitivity analysis of nonlinear models},}\
  }\href {\doibase https://doi.org/10.1016/0951-8320(96)00002-6} {\bibfield
  {journal} {\bibinfo  {journal} {Reliability Engineering and System Safety}\
  }\textbf {\bibinfo {volume} {52}},\ \bibinfo {pages} {1 -- 17} (\bibinfo
  {year} {1996})}\BibitemShut {NoStop}%
\bibitem [{\citenamefont {Gelman}\ \emph {et~al.}(2013)\citenamefont {Gelman},
  \citenamefont {Carlin}, \citenamefont {Stern}, \citenamefont {Dunson},
  \citenamefont {Vehtari},\ and\ \citenamefont {Rubin}}]{gelman}%
  \BibitemOpen
  \bibfield  {author} {\bibinfo {author} {\bibfnamefont {A}~\bibnamefont
  {Gelman}}, \bibinfo {author} {\bibfnamefont {J~B}\ \bibnamefont {Carlin}},
  \bibinfo {author} {\bibfnamefont {H~S}\ \bibnamefont {Stern}}, \bibinfo
  {author} {\bibfnamefont {D~B}\ \bibnamefont {Dunson}}, \bibinfo {author}
  {\bibfnamefont {A}~\bibnamefont {Vehtari}}, \ and\ \bibinfo {author}
  {\bibfnamefont {D~B}\ \bibnamefont {Rubin}},\ }\href@noop {} {\emph {\bibinfo
  {title} {Bayesian Data Analysis, Third Edition}}},\ Chapman Hall/CRC Texts in
  Statistical Science\ (\bibinfo  {publisher} {Taylor Francis},\ \bibinfo
  {year} {2013})\BibitemShut {NoStop}%
\bibitem [{\citenamefont {McDonnell}\ \emph {et~al.}(2015)\citenamefont
  {McDonnell}, \citenamefont {Schunck}, \citenamefont {Higdon}, \citenamefont
  {Sarich}, \citenamefont {Wild},\ and\ \citenamefont
  {Nazarewicz}}]{McDonnell2015}%
  \BibitemOpen
  \bibfield  {author} {\bibinfo {author} {\bibfnamefont {J.~D.}\ \bibnamefont
  {McDonnell}}, \bibinfo {author} {\bibfnamefont {N.}~\bibnamefont {Schunck}},
  \bibinfo {author} {\bibfnamefont {D.}~\bibnamefont {Higdon}}, \bibinfo
  {author} {\bibfnamefont {J.}~\bibnamefont {Sarich}}, \bibinfo {author}
  {\bibfnamefont {S.~M.}\ \bibnamefont {Wild}}, \ and\ \bibinfo {author}
  {\bibfnamefont {W.}~\bibnamefont {Nazarewicz}},\ }\bibfield  {title}
  {\enquote {\bibinfo {title} {Uncertainty quantification for nuclear density
  functional theory and information content of new measurements},}\ }\href
  {\doibase 10.1103/PhysRevLett.114.122501} {\bibfield  {journal} {\bibinfo
  {journal} {Phys. Rev. Lett.}\ }\textbf {\bibinfo {volume} {114}},\ \bibinfo
  {pages} {122501} (\bibinfo {year} {2015})}\BibitemShut {NoStop}%
\bibitem [{\citenamefont {Neufcourt}\ \emph {et~al.}(2019)\citenamefont
  {Neufcourt}, \citenamefont {Cao}, \citenamefont {Nazarewicz}, \citenamefont
  {Olsen},\ and\ \citenamefont {Viens}}]{Neufcourt2019}%
  \BibitemOpen
  \bibfield  {author} {\bibinfo {author} {\bibfnamefont {L\'eo}\ \bibnamefont
  {Neufcourt}}, \bibinfo {author} {\bibfnamefont {Yuchen}\ \bibnamefont {Cao}},
  \bibinfo {author} {\bibfnamefont {Witold}\ \bibnamefont {Nazarewicz}},
  \bibinfo {author} {\bibfnamefont {Erik}\ \bibnamefont {Olsen}}, \ and\
  \bibinfo {author} {\bibfnamefont {Frederi}\ \bibnamefont {Viens}},\
  }\bibfield  {title} {\enquote {\bibinfo {title} {Neutron drip line in the ca
  region from bayesian model averaging},}\ }\href {\doibase
  10.1103/PhysRevLett.122.062502} {\bibfield  {journal} {\bibinfo  {journal}
  {Phys. Rev. Lett.}\ }\textbf {\bibinfo {volume} {122}},\ \bibinfo {pages}
  {062502} (\bibinfo {year} {2019})}\BibitemShut {NoStop}%
\bibitem [{\citenamefont {Vernon}\ \emph {et~al.}(2010)\citenamefont {Vernon},
  \citenamefont {Goldstein},\ and\ \citenamefont {Bower}}]{Vernon2010}%
  \BibitemOpen
  \bibfield  {author} {\bibinfo {author} {\bibfnamefont {Ian}\ \bibnamefont
  {Vernon}}, \bibinfo {author} {\bibfnamefont {Michael}\ \bibnamefont
  {Goldstein}}, \ and\ \bibinfo {author} {\bibfnamefont {Richard~G}\
  \bibnamefont {Bower}},\ }\bibfield  {title} {\enquote {\bibinfo {title}
  {{Galaxy formation: a Bayesian uncertainty analysis}},}\ }\href {\doibase
  10.1214/10-BA524} {\bibfield  {journal} {\bibinfo  {journal} {Bayesian
  Analysis}\ }\textbf {\bibinfo {volume} {5}},\ \bibinfo {pages} {619--669}
  (\bibinfo {year} {2010})}\BibitemShut {NoStop}%
\bibitem [{\citenamefont {{Carlsson}}\ \emph {et~al.}(2015)\citenamefont
  {{Carlsson}}, \citenamefont {{Ekstr{\"o}m}}, \citenamefont {{Forss{\'e}n}},
  \citenamefont {{Fahlin Str{\"o}mberg}}, \citenamefont {{Jansen}},
  \citenamefont {{Lilja}}, \citenamefont {{Lindby}}, \citenamefont
  {{Mattsson}},\ and\ \citenamefont {{Wendt}}}]{carlsson2015}%
  \BibitemOpen
  \bibfield  {author} {\bibinfo {author} {\bibfnamefont {B.~D.}\ \bibnamefont
  {{Carlsson}}}, \bibinfo {author} {\bibfnamefont {A.}~\bibnamefont
  {{Ekstr{\"o}m}}}, \bibinfo {author} {\bibfnamefont {C.}~\bibnamefont
  {{Forss{\'e}n}}}, \bibinfo {author} {\bibfnamefont {D.}~\bibnamefont {{Fahlin
  Str{\"o}mberg}}}, \bibinfo {author} {\bibfnamefont {G.~R.}\ \bibnamefont
  {{Jansen}}}, \bibinfo {author} {\bibfnamefont {O.}~\bibnamefont {{Lilja}}},
  \bibinfo {author} {\bibfnamefont {M.}~\bibnamefont {{Lindby}}}, \bibinfo
  {author} {\bibfnamefont {B.~A.}\ \bibnamefont {{Mattsson}}}, \ and\ \bibinfo
  {author} {\bibfnamefont {K.~A.}\ \bibnamefont {{Wendt}}},\ }\bibfield
  {title} {\enquote {\bibinfo {title} {{Uncertainty analysis and order-by-order
  optimization of chiral nuclear interactions}},}\ }\href
  {http://adsabs.harvard.edu/abs/2015arXiv150602466C} {\bibfield  {journal}
  {\bibinfo  {journal} {ArXiv e-prints}\ } (\bibinfo {year} {2015})},\ \Eprint
  {http://arxiv.org/abs/1506.02466} {arXiv:1506.02466 [nucl-th]} \BibitemShut
  {NoStop}%
\bibitem [{\citenamefont {Contessi}\ \emph {et~al.}(2017)\citenamefont
  {Contessi}, \citenamefont {Lovato}, \citenamefont {Pederiva}, \citenamefont
  {Roggero}, \citenamefont {Kirscher},\ and\ \citenamefont {van
  Kolck}}]{contessi2017}%
  \BibitemOpen
  \bibfield  {author} {\bibinfo {author} {\bibfnamefont {L.}~\bibnamefont
  {Contessi}}, \bibinfo {author} {\bibfnamefont {A.}~\bibnamefont {Lovato}},
  \bibinfo {author} {\bibfnamefont {F.}~\bibnamefont {Pederiva}}, \bibinfo
  {author} {\bibfnamefont {A.}~\bibnamefont {Roggero}}, \bibinfo {author}
  {\bibfnamefont {J.}~\bibnamefont {Kirscher}}, \ and\ \bibinfo {author}
  {\bibfnamefont {U.}~\bibnamefont {van Kolck}},\ }\bibfield  {title} {\enquote
  {\bibinfo {title} {Ground-state properties of 4he and 16o extrapolated from
  lattice qcd with pionless eft},}\ }\href {\doibase
  https://doi.org/10.1016/j.physletb.2017.07.048} {\bibfield  {journal}
  {\bibinfo  {journal} {Physics Letters B}\ }\textbf {\bibinfo {volume}
  {772}},\ \bibinfo {pages} {839 -- 848} (\bibinfo {year} {2017})}\BibitemShut
  {NoStop}%
\bibitem [{\citenamefont {Bansal}\ \emph {et~al.}(2018)\citenamefont {Bansal},
  \citenamefont {Binder}, \citenamefont {Ekstr\"om}, \citenamefont {Hagen},
  \citenamefont {Jansen},\ and\ \citenamefont {Papenbrock}}]{bansal2018}%
  \BibitemOpen
  \bibfield  {author} {\bibinfo {author} {\bibfnamefont {A.}~\bibnamefont
  {Bansal}}, \bibinfo {author} {\bibfnamefont {S.}~\bibnamefont {Binder}},
  \bibinfo {author} {\bibfnamefont {A.}~\bibnamefont {Ekstr\"om}}, \bibinfo
  {author} {\bibfnamefont {G.}~\bibnamefont {Hagen}}, \bibinfo {author}
  {\bibfnamefont {G.~R.}\ \bibnamefont {Jansen}}, \ and\ \bibinfo {author}
  {\bibfnamefont {T.}~\bibnamefont {Papenbrock}},\ }\bibfield  {title}
  {\enquote {\bibinfo {title} {Pion-less effective field theory for atomic
  nuclei and lattice nuclei},}\ }\href {\doibase 10.1103/PhysRevC.98.054301}
  {\bibfield  {journal} {\bibinfo  {journal} {Phys. Rev. C}\ }\textbf {\bibinfo
  {volume} {98}},\ \bibinfo {pages} {054301} (\bibinfo {year}
  {2018})}\BibitemShut {NoStop}%
\end{thebibliography}%
\end{document}